\documentclass{aa}

\usepackage{graphicx}
\usepackage{txfonts}
\usepackage{url}
\usepackage{wasysym}
\usepackage{longtable}
\usepackage{array}
\usepackage{tikz}
\usepackage[pdftitle={Transit least-squares survey IV. Earth-like transiting planets expected from the PLATO mission}, colorlinks = true, breaklinks = true, citecolor = blue, linkcolor = blue, urlcolor = blue, pdfauthor = {Ren\'{e} Heller}]{hyperref}
\usepackage{esvect}  

\begin{document} 

\title{Transit least-squares survey}
\subtitle{IV. Earth-like transiting planets expected from the PLATO mission}
\titlerunning{Transit least-squares survey -- IV. Earth-like planets from PLATO}

\author{
  Ren{\'e} Heller\inst{1}
  \and
  Jan-Vincent Harre\inst{2}
  \and
  R{\'e}za Samadi\inst{3}
}

\institute{
Max-Planck-Institut f\"ur Sonnensystemforschung, Justus-von-Liebig-Weg 3, 37077 G\"ottingen, Germany; \href{mailto:heller@mps.mpg.de}{heller@mps.mpg.de}
\and
Institute of Planetary Research, German Aerospace Center, Rutherfordstrasse 2, 12489 Berlin, Germany; \href{mailto:jan-vincent.harre@dlr.de}{jan-vincent.harre@dlr.de}
\and
LESIA, Observatoire de Paris, Universit{\'e} PSL, CNRS, Sorbonne Universit{\'e}, Universit{\'e} Paris Diderot, Sorbonne Paris Cit{\'e}, 5 place Jules Janssen, 92195 Meudon, France; \href{mailto:reza.samadi@obspm.fr}{reza.samadi@obspm.fr}
}

   \date{Received 28 June 2021; Accepted 20 June 2022}

 
\abstract{
In its long-duration observation phase, the PLATO satellite (scheduled for launch in 2026) will observe two independent, non-overlapping fields, nominally one in the northern hemisphere and one in the southern hemisphere, for a total of four years. The exact duration of each pointing will be determined two years before launch. Previous estimates of PLATO's yield of Earth-sized planets in the habitable zones (HZs) around solar-type stars ranged between 6 and 280. We use the PLATO Solar-like Light curve Simulator ({\tt PSLS}) to simulate light curves with transiting planets around bright ($m_V~{\leq}~11$) Sun-like stars at a cadence of 25\,s, roughly representative of the $>15,000$ targets in PLATO's high-priority P1 sample (mostly F5-K7 dwarfs and subdwarfs). Our study includes light curves generated from synchronous observations of 6, 12, 18, and 24 of PLATO's 12\,cm aperture cameras over both 2\,yr and 3\,yr of continuous observations. Automated detrending is done with the {\tt W{\={o}}tan} software, and post-detrending transit detection is performed with the transit least-squares ({\tt TLS}) algorithm. Light curves combined from 24 cameras yield true positive rates (TPRs) near unity for planets ${\geq}1.2\,R_\oplus$ with two transits. If a third transit is in the light curve, planets as small as $1\,R_\oplus$ are recovered with TPR\,\,$\sim100$\,\%. We scale the TPRs with the expected number of stars in the P1 sample and with modern estimates of the exoplanet occurrence rates and predict the detection of planets with $0.5\,R_\oplus\,\leq\,R_{\rm p}\,\leq\,1.5\,R_\oplus$ in the HZs around F5-K7 dwarf stars. For the long-duration observation phase  (2\,yr~+~2\,yr) strategy we predict 11--34 detections, and for the (3\,yr~+~1\,yr) strategy we predict 8--25 discoveries. These estimates neglect exoplanets with monotransits, serendipitous detections in stellar samples P2--P5, a dedicated removal of systematic effects, and a possible bias of the P1 sample toward brighter stars and high camera coverage due to noise requirements. As an opposite effect, Earth-sized planets might typically exhibit transits around P1 sample stars shallower than we have assumed since the P1 sample will be skewed toward spectral types earlier than the Sun-like stars assumed in our simulations. Moreover, our study of the effects of stellar variability on shallow transits of Earth-like planets illustrates that our estimates of PLATO's planet yield, which we derive using a photometrically quiet star similar to the Sun, must be seen as upper limits. In conclusion, PLATO's detection of about a dozen Earth-sized planets in the HZs around solar-type stars will mean a major contribution to this as yet poorly sampled part of the exoplanet parameter space with Earth-like planets.}


   \maketitle
%

\section{Introduction}


\begin{figure*}[h]
    \centering
    \includegraphics[width=.98\linewidth]{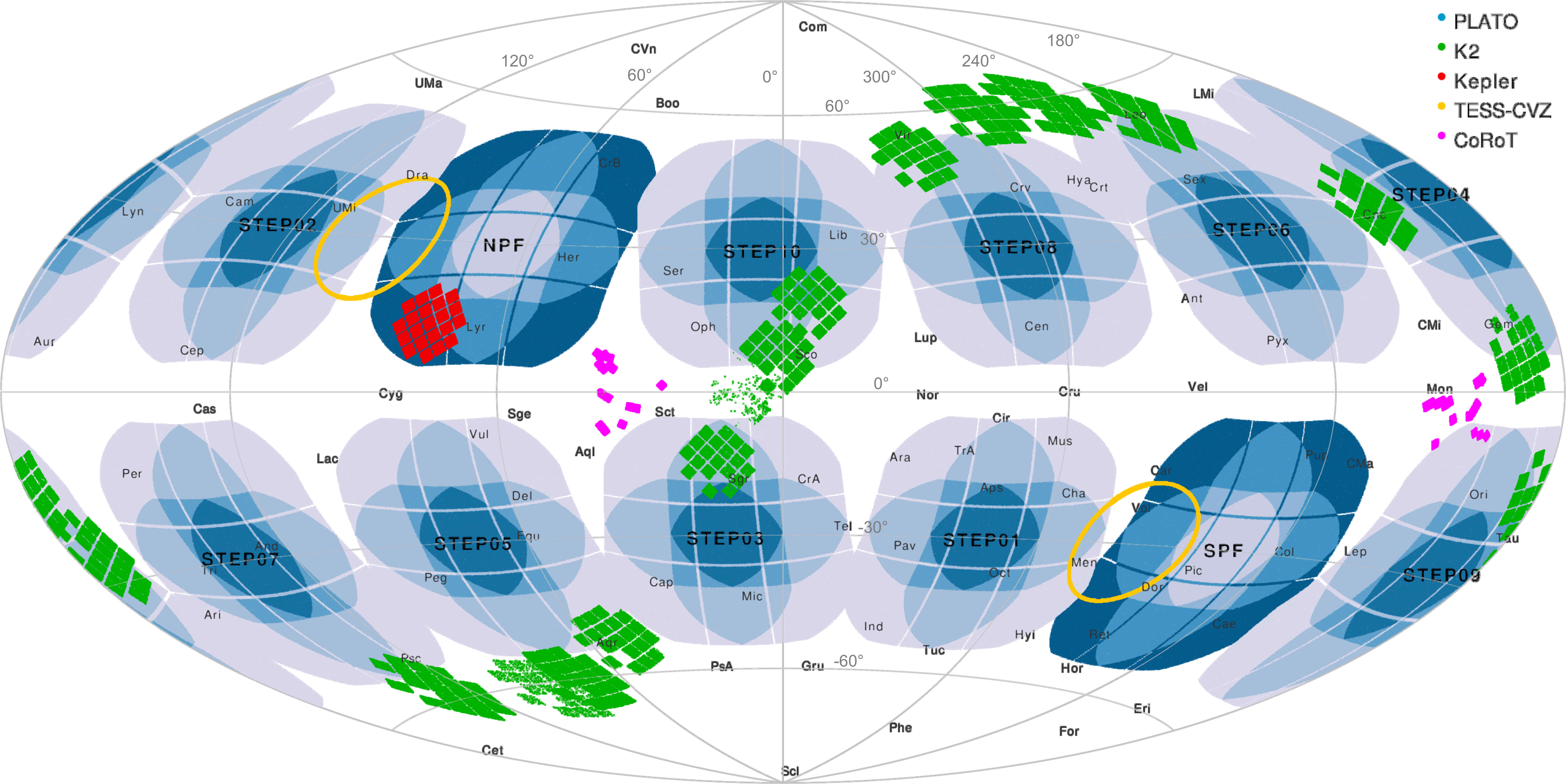}
    \caption{Provisional pointing of PLATO's northern (NPF) and southern (SPF) long-duration phase fields in an all-sky Aitoff projection using Galactic coordinates \citep[an updated version is available in][]{2022A&A...658A..31N}. The centers of the NPF and SPF are covered by 24 cameras (light blue). Increasingly darker blue tones refer to a coverage by 18, 12, and 6 cameras, respectively. Provisional step-and-stare field pointings (STEP01 -- STEP10) use dark blue tones for an overlap of 24 cameras and lighter blue tones for 18, 12, and 6 cameras, respectively. The fields of CoRoT (pink), Kepler (red), K2 (green), and the TESS continuous viewing zone (yellow) are also shown.}
    \label{fig:PLATO_fov}
\end{figure*}

The science of extrasolar planets is driven to a large extent by transiting planets, that is, planets passing in front of their host stars once every orbit as seen from Earth \citep{2000ApJ...529L..45C}. In fact, most of the planets known beyond the Solar System have been found via the transit method in the long-term stellar light curves from space-based missions, starting with 37 planets and brown dwarfs from the CoRoT mission \citep[2009-2013;][]{2009A&A...506..411A,2018A&A...619A..97D}, over 2700 exoplanets and about 2100 candidates yet to be validated\footnote{NASA Exoplanet Archive, \href{https://exoplanetarchive.ipac.caltech.edu}{https://exoplanetarchive.ipac.caltech.edu}, as of 20 June 2022} from the Kepler primary mission \citep[2009-2013;][]{2010Sci...327..977B}, more than 500 transiting planets and almost 1000 candidates\footnotemark[\value{footnote}] from the extended Kepler mission, called K2 \citep[2014-2018;][]{2014PASP..126..398H}, and the addition of over 200 exoplanets\footnotemark[\value{footnote}] discovered with the ongoing TESS mission \citep[since 2018;][]{2015JATIS...1a4003R}.

Despite the rapid increase in the number of known planets by about three orders of magnitude within three decades \citep{2019arXiv191112114H}, it remains extremely challenging to find Earth-sized planets in Earth-like orbits around Sun-like stars. As a result, it is hard to constrain their occurrence rates, even with four years of continuous observations from the Kepler primary mission \citep{2019ApJ...887..157M,2019AJ....158..109H,2020AJ....159..248K,2021AJ....161...36B}. The key challenges for Kepler were the small number of transits of these planets (Kepler required three), a systematic effect in the Kepler data known as the rolling-band effect \citep{2016ksci.rept....1V,2018ApJS..235...38T,2021AJ....161...36B}, and the fact that stars are more variable than predicted via the scaling of the solar power spectrum \citep{2011ApJS..197....6G,2015AJ....150..133G}. Even detections of super-Earth-sized planets in one-year orbits around Sun-like stars such as Kepler-452\,b \citep{2015AJ....150...56J} and KOI-456.04 \citep{2020A&A...638A..10H} have been challenging due to these caveats \citep{2018AJ....155..210M}.

The science objective of the PLATO mission \citep{2014ExA....38..249R}, with an expected launch in December 2026, is to find terrestrial planets in the habitable zones \citep[HZs;][]{1993Icar..101..108K} of solar-like stars and to determine their bulk properties\footnote{PLATO Definition Study Report, \href{https://sci.esa.int/s/8rPyPew}{https://sci.esa.int/s/8rPyPew}}. To achieve this goal, PLATO's 24 normal cameras\footnote{The 24 ``normal'' cameras will observe stars with apparent visual magnitudes $m_V \gtrsim 8$ in white light with a cadence of 25\,s. Two additional ``fast'' cameras will observe very bright stars with $4 \lesssim m_V \lesssim 8$ in two different color filters (one in a blue filter, the other in a red filter) with a cadence of 2.5\,s. The fast cameras will be used for color refinements of astrophysical processes and fine guidance.} will observe over $15,000$ bright ($m_V \leq 11$) dwarfs and subgiants of spectral types F5-K7 with a precision of 50\,ppm in 1\,hr of integration \citep{2021A&A...653A..98M}. This subset of the hundreds of thousands of stars that PLATO will observe in total is referred to as the P1 stellar sample, and it will be given the highest priority of all the samples. Over $5000$ stars with $m_V \leq 11$ will have their bulk densities measured using asteroseismology. The P1 sample will be observed during the long-duration observation phase, which will take a total of four years, and be compiled from the northern PLATO field (NPF) and the southern PLATO field (SPF). Figure~\ref{fig:PLATO_fov} shows the distribution of the PLATO fields that include the preliminary NPF and SPF on the celestial plane. A more up-to-date definition of these fields, which are referred to as Long-duration Observation Phase North 1 (LOPN1) and Long-duration Observation Phase South 1 (LOPS1), is given in \citep{2022A&A...658A..31N}.

The Gaia mission has already provided highly accurate parallax and optical photometry measurements of essentially all main-sequence stars with $m_V \lesssim 20$ \citep{2016A&A...595A...1G,2021A&A...649A...1G}. In combination with effective temperature measurements from PLATO's ground-based high-resolution spectroscopy follow-up, stellar radii will be determined with an accuracy of 1--2\,\%, masses will be inferred with an accuracy of $10$\,\%, and ages will be estimated using stellar evolution models with an accuracy of 10\,\% for over 5\,000 F5-K7 stars.

The PLATO Definition Study Report predicted PLATO's yield of Earth-sized transiting planets in the HZ around Sun-like stars to range between 6 and 280. The details depend on PLATO's observing strategy, that is, whether it observes its two independent long-duration observation phase fields for 2\,yr each (the 2\,yr + 2\,yr scenario) or one field for 3\,yr and the other one for 1\,yr (the 3\,yr + 1\,yr scenario). These yield predictions were derived with exoplanet occurrence rates that were still very uncertain at the time \citep{2013ApJ...766...81F}. The PLATO Definition Study Report considered an occurrence rate of planets in the HZ with radii smaller than two Earth radii ($R_{\rm p}~<~2\,R_\oplus$) of between 2\,\% and 100\,\%, with a nominal value of 40\,\%. The most recent and comprehensive statistical analysis of the completeness and reliability of the Kepler survey by \citet{2021AJ....161...36B} suggests that the occurrence rate of planets with $0.5\,R_\oplus~\leq~R_{\rm p}~\leq~1.5\,R_\oplus$ and orbital periods between 340\,d and 760\,d \citep[the conservative HZ;][]{2014ApJ...787L..29K} around G and K dwarf stars is closer to 60\,\%. Moreover, previous yield predictions for PLATO were necessarily based on analytical estimates of the expected signal-to-noise ratio (S/N) of the transits and taking the activity of Sun-like stars observed with Kepler into account.

With the launch date of the PLATO mission approaching, great progress has been made in the simulation of light curves that can be expected from PLATO. The {\tt PLATOSim} software \citep{2014A&A...566A..92M} provides end-to-end simulations from the charge-coupled device (CCD) level to the final light curve product. The PLATO Solar-like Light curve Simulator ({\tt PSLS}), following a much more pragmatic and computationally efficient approach, allows simulations of a large number of PLATO-like light curves with realistic treatment of the main sources of astrophysical and systematic noise \citep{Samadi_2019}. As for transit detection, the {\tt W{\={o}}tan} detrending software for stellar light curves has been optimized with a particular focus on the preservation of exoplanet transits \citep{wotan_Hippke_2019}. This optimization was achieved based on extended benchmarking tests invoking detrending and transit searches in light curves from the Kepler primary mission, K2, and TESS. Finally, the release of the transit least-squares ({\tt TLS}) software provides enhanced detection sensitivity for small planets over the traditional box least-squares ({\tt BLS}) algorithm \citep{kovacs2002box}. As shown by \citet{Hippke_Heller_2019_tls}, for a parameterization of {\tt TLS} and {\tt BLS} that produces a false positive rate (FPR) of 1\,\% in a search for transits of Earth-like planets around Sun-like stars, the true positive rate (TPR) of {\tt TLS} is about 93\,\%, while it is roughly 76\,\% for {\tt BLS}. This sensitivity gain of {\tt TLS} over {\tt BLS} will be vital for PLATO's discovery of Earth-sized planets in the HZs around Sun-like stars.

The key objectives of this paper are to update PLATO's expected yield of Earth-like planets in the HZs around Sun-like stars and to test if currently available transit search software would be sufficient to find these planets in PLATO data and enable PLATO to achieve its science objectives.

\section{Methods}

The principal approach of our study is as follows. We used {\tt PSLS} to simulate a sufficiently large number (${\sim}10,000$) of PLATO-like light curves for Sun-like stars, including stellar variability and systematic effects, some with transits of Earth-like planets, some without any transits. Then we used {\tt W{\={o}}tan} to automatically detrend the light curves from any stellar and systematic variability while preserving any possible transit signals. To infer the detectability of the injected transits as well as the corresponding false alarm probability, we searched for the transits using {\tt TLS}. A comparison of the injected signals with our results from the recovery tests then yielded the true and FPRs, which we scaled with literature values for the planet occurrence rates to predict the number of Earth-like planets to be detected with PLATO.

\subsection{PLATO Solar-like Light Curve Simulator}

Our analysis starts with the generation of synthetic PLATO-like light curves with the publicly available\footnote{\href{https://sites.lesia.obspm.fr/psls/}{https://sites.lesia.obspm.fr/psls}} {\tt PSLS} (v1.3) Python software \citep{Samadi_2019}. {\tt PSLS} simulates Poisson noise characteristic for a given stellar magnitude, instrumental effects, systematic errors, and stellar variability.

In a first step, {\tt PSLS} reads the input YAML file to simulate the stellar signal in Fourier space. Then random noise is added \citep{1990ApJ...364..699A} to mimic the stochastic behavior of the signal and finally the signal is transformed back into the time domain to generate the light curve.

The stellar oscillation spectrum is computed as a sum of resolved and unresolved differential modes, which are modeled with specific Lorentzian profiles in the power spectral density space \citep{2001ESASP.464..411B,Samadi_2019}. The mode frequencies, mode heights, and mode line widths for main-sequence and subgiant stars are precomputed with the Aarhus adiabatic pulsation package ({\tt ADIPLS}) \citep{Christensen_Dalsgaard_2008}. Stellar activity phenomena such as magnetic activity (star spots), p-mode oscillations, and granulation lead to time-dependent variability for the disk-integrated flux of a solar-like star. The stellar activity component is simulated in {\tt PSLS} with a Lorentzian profile in frequency space, and it includes an amplitude ($\sigma_A$, subscript ``A'' referring to activity) and a characteristic timescale ($\tau_A$), both of which can be adjusted in {\tt PSLS} \citep{Samadi_2019}. Stellar granulation, caused by convection currents of plasma within the star's convective zone, occurs on a scale from granules the size of ${\sim}0.5\,$Mm (${\sim}0.08\,R_\oplus$) to supergranules with diameters of ${\sim}16\,$Mm (${\sim}2.5\,R_\oplus$), all of which  appear stochastically over time \citep{2020MNRAS.493.5489M}. Granulation is simulated in {\tt PSLS} using the two pseudo-Lorentzian functions with characteristic timescales $\tau_{1,2}$ \citep{Samadi_2019}.

Systematic errors of the PLATO instrument are simulated in {\tt PSLS} using the Plato Image Simulator ({\tt PIS}), developed at the Laboratoire d'{\'e}tudes spatiales et d'instrumentation en astrophysique (LESIA) at the Observatoire de Paris. {\tt PIS} models different sources of noise and other perturbations, like photon noise, readout noise, smearing, long-term drift, satellite jitter, background signal, intra-pixel response nonuniformity, pixel response nonuniformity, digital saturation \citep{janesick2001scientific,2014A&A...566A..92M}, charge diffusion \citep{1999PASP..111.1434L}, and charge transfer inefficiency \citep{2013MNRAS.430.3078S}. In our simulations we used the beginning-of-life setup tables for systematics, where charge transfer inefficiency is not included in simulating systematics. We also turned off jitter in {\tt PIS} as it demands substantial amounts of CPU time. Furthermore, according to the PLATO Study Definition Report, the pointing errors are expected to be sufficiently low that they will be negligible to the overall noise \citep[see also][]{2019A&A...627A..71M}. Light curves for the P1 sample will not be extracted from aperture masks but from the downloaded imagettes using the point-spread function fitting method. The resulting systematic effects, including jumps in the light curves at the quarterly repositioning of the stars on the PLATO CCDs after satellite rotation, are properly taken into account in {\tt PSLS}.

Finally, planetary transits can be automatically simulated in {\tt PSLS} data using the \citet{2002ApJ...580L.171M} model. The actual implementation in {\tt PSLS} is based on the Python code of Ian Crossfield\footnote{\href{https://www.mit.edu/~iancross/python/}{https://www.mit.edu/~iancross/python}}. {\tt PSLS} users can specify the transit parameters, including the planet radius ($R_{\rm p}$), the orbital period ($P$), the planet semimajor axis ($a$), and the orbital inclination. Transits assume a quadratic limb darkening law as per \citet{2002ApJ...580L.171M}, and the two quadratic limb darkening coefficients of the star can be manually adjusted.

For all our simulations, we chose a solar-type star with solar radius and mass and with Sun-like stellar activity to represent the expected 15,000 to 20,000 F5 - K7 stars in the P1 sample. As for the amplitude and characteristic timescale of stellar photometric activity, we assumed activity parameters close to those of 16\,Cyg\,B, that is, $\sigma_{\rm A}=20$\,ppm and $\tau_{\rm A}=0.27$\,d. This is similar to the values used in the default parameterization of the {\tt PSLS} YAML input file, as described in \citet[][Appendix~A therein]{Samadi_2019}. 16\,Cyg\,B is a solar-type oscillating main-sequence star, for which stellar activity, asteroseismology, and rotation periods have been well constrained using Kepler data \citep{2015MNRAS.446.2959D}. In Appendix~\ref{app:activity} we examine the effects of different levels of stellar variability on transit detection with PLATO.

The resulting {\tt PSLS} light curves are derived as averages from 6, 12, 18, or 24 light curves (corresponding to the number of cameras) and have a cadence of 25\,s, representative of the data that will be extracted for the P1 sample. For reference, a {\tt PSLS} light curve worth of 2\,yr (3\,yr) of simulated observations contains about 2.5 (3.7) million data points.

\subsection{Light curve detrending with {\tt W{\={o}}tan}}

{\tt W{\={o}}tan} is a publicly available\footnote{\href{https://github.com/hippke/wotan}{https://github.com/hippke/wotan}} Python software for the detrending of stellar light curves under optimized preservation of exoplanet transit signatures \citep{wotan_Hippke_2019}. The key value of {\tt W{\={o}}tan} is in its removal of instrumental and stellar variability from light curves to prepare them for transit searches. {\tt W{\={o}}tan} features many different detrending methods. Among all these detrending filters, \citet{wotan_Hippke_2019} identified the biweight method as the optimal choice in most cases.

We thus use the biweight method in this work with a window size of three times the expected maximum transit duration, as suggested by \citet{wotan_Hippke_2019}, to preserve the transit signatures while removing any other type of variability from the simulated light curves from {\tt PSLS}. It is worth noting that while we restrict our search to transits of planets in Earth-like orbits and, hence, Earth-like transit durations, in an unrestricted search for transits of different durations the window size of the detrending algorithm would need to be adapted accordingly. {\tt W{\={o}}tan}'s treatment of jumps in the {\tt PSLS} light curves is described in Sect.~\ref{sec:limitations}.

\subsection{Transit Least-Squares}

The transit search is executed with the publicly available\footnote{\href{https://github.com/hippke/tls}{https://github.com/hippke/tls}} {\tt TLS} algorithm \citep{Hippke_Heller_2019_tls}. {\tt TLS} is optimized for finding small planets. The principal sensitivity gain over the {\tt BLS} algorithm \citep{kovacs2002box} is in {\tt TLS}'s accurate modeling of the planetary ingress and egress and stellar limb-darkening. The template of {\tt TLS}'s transit search function is parameterized by two limb darkening coefficients required to feed the analytical solution for the light curve with a quadratic limb darkening law \citep{2002ApJ...580L.171M}.

{\tt TLS} finds the minimum $\chi^2$ value for a range of trial orbital periods ($P$), transit times $t_0$, and transit durations $d$. The orbital eccentricity is assumed to be zero, but the transit sequence of planets in highly eccentric orbits are typically found with {\tt TLS} as long as the transits are periodic. {\tt TLS} does not fit for transit timing variations (TTVs), but transits are usually recovered as long as the TTV amplitude is smaller than the transit duration.

The key search metric for {\tt TLS} is the signal detection efficiency (SDE), which is a measure for the significance of the $\chi^2$ minimum compared to the surrounding $\chi^2$ landscape as a function of the period. \citet{Hippke_Heller_2019_tls}, using simulated transits of Earth-like planets around Sun-like stars in synthetic light curves with a white noise component of 110\,ppm per 30\,min integration (corresponding to a photometrically quiet $m_V=12$ star observed with Kepler), found that a detection threshold of ${\rm SDE}~\geq~9$ results in an FPR $<~10^{-4}$ for {\tt TLS}. At the same time, various studies have demonstrated that this threshold is sufficiently low for {\tt TLS} to detect Earth-sized planets around low-mass stars in Kepler \citep{2019A&A...625A..31H,2019A&A...627A..66H} and TESS \citep{2021Sci...371.1038T,2021MNRAS.502.2845R,2021AJ....161..247F} data, super-Earths around solar-type stars \citep{2020A&A...638A..10H}, and possibly even Mars-sized planets in close orbits around subdwarfs from Kepler \citep{2021A&A...650A.205V}. Hence, we use ${\rm SDE}=9$ in this work as well. Moreover, we require ${\rm S/N}>7$ for any candidate signal to count as a detection. This threshold has been shown to yield one statistical false alarm in a sample of 100,000 photometrically quiet stars from the Kepler mission with normally distributed noise \citep[][Sect.~2.3 therein]{2002ApJ...564..495J}. {\tt TLS} calculates the S/N as $(\delta/\sigma_{\rm o})\,n^{1/2}$, where $\delta$ is the mean depth of the trial transit, $\sigma_{\rm o}$ is the standard deviation of the out-of-transit data, and $n$ is the number of in-transit data points \citep{2006MNRAS.373..231P}. We do not use any binning of the light curves in {\tt TLS} because our focus is completeness rather than computational speed. We mask out any residual jumps in the light curve after detrending using the {\tt transit\_mask} parameter in {\tt TLS}.

\begin{figure}
    \centering
    \includegraphics[width=\linewidth]{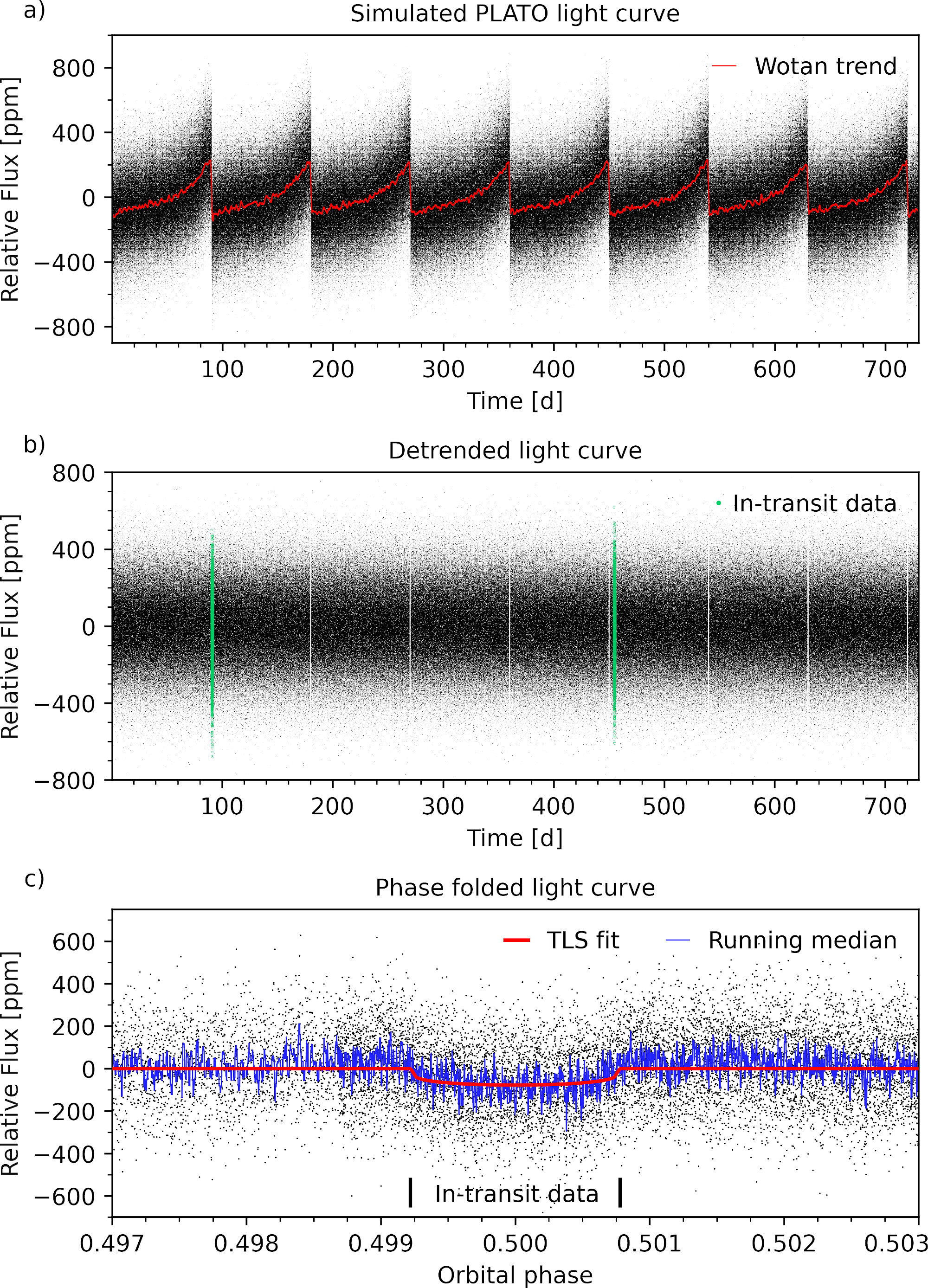}
    \caption{Illustration of our transit injection and retrieval experiment. Simulations include two transits of an Earth-sized planet with an orbital period of 364\,d in a 2\,yr light curve of an $m_V = 9$ Sun-like star. (a) PLATO-like light curve generated with {\tt PSLS}. The red line shows the trend computed with {\tt W{\={o}}tan}. (b) Light curve after detrending with {\tt W{\={o}}tan}. The simulated transits occur at about 91\,d and 455\,d (highlighted with green points), that is, near quarterly reorientations of the spacecraft. This example was deliberately chosen to illustrate that {\tt W{\={o}}tan} and {\tt TLS} can master non-well-behaved systems. (c) Phase-folded light curve of the transit detected with {\tt TLS}. The best model fit is shown with a red line. The 21-bin walking median (ten data points symmetrically around each data point) is shown with a blue line.}
    \label{fig:method_overview}
\end{figure}

In Fig.~\ref{fig:method_overview} we summarize the sequence of steps executed with {\tt PSLS}, {\tt W{\={o}}tan}, and {\tt TLS}. In panel (a), trends are dominantly caused by spacecraft drift, inducing changes of the camera pointings and a motion of the target star across CCD pixels with different quantum sensitivities. Jumps result from satellite rotation and the resulting repositioning of the star on the PLATO CCDs at the beginning of each PLATO quarter. Panel (b) shows the detrended light curve, which we obtain by dividing the simulated light curve from {\tt PSLS} (black points in (a)) by the trend computed with {\tt W{\={o}}tan} (red line in (a)). Panel (c) presents the phase-folded light curve of the transit detected with {\tt TLS}, including the phase-folded {\tt PSLS} data (black dots), the {\tt TLS} best fitting transit model (red line), and the 21-bin walking median of the data (blue line). In this example, {\tt TLS} produces SDE~=~7.3 and S/N~=~24.5. We point out that the purpose of {\tt TLS} is not exoplanet characterization but exoplanet detection. A Markov chain Monte Carlo fitting procedure that takes into account the stellar mass and radius as well as the best-fit values for $t_0$, $P$, and $d$ from {\tt TLS} as priors \citep[as in][]{2020A&A...638A..10H} can improve the system parameter estimates substantially.

\subsection{Exoplanet occurrence rates and transit probability}
\label{sec:occurrence}

\citet{Petigura2013} showed that the occurrence rate of exoplanets with $2\,R_\oplus<R_{\rm p}<4\,R_\oplus$ and $200\,{\rm d}<P<400$\,d around Sun-like stars is  $5\,(\pm\,2.1)\,\%$. The occurrence rate of smaller planets in this period range was unconstrained. A thorough treatment of Kepler's false positive transit events by \citet{2021AJ....161...36B} resulted in an occurrence rate of planets with radii $0.5\,R_\oplus~<~R_{\rm p}~<~1.5\,R_\oplus$ and inside the conservative HZs around GK stars close to $\sim60\,\%$, details depending on the parameterization of the HZ as a function of the stellar properties and on the reliability of exoplanet detections against systematic false positives \citep{2020AJ....159..279B}. For our purpose of estimating PLATO's Earth-like planet yield, we assume a 37\,\% occurrence rate in the conservative scenario and 88\,\% in the optimistic scenario, referring to stellar population Model~1 and hab2 stars in Table~3 of \citet{2021AJ....161...36B}.

PLATO's P1 sample contains at least 15,000 and up to 20,000 dwarf and subgiant bright ($m_V~{\leq}~11$) stars with spectral types F5-K7. Detailed histograms for the mass, radius, effective temperature, distance, and Gaia magnitude distributions of the PLATO stellar sample have recently been published by \citet{2021A&A...653A..98M}, but were not available before the end of our simulations. We consider 15,000 stars for our conservative scenario and 20,000 stars for our optimistic scenario. We assume that the P1 stars will be equally distributed over the NPF and SPF during the long-duration observation phase (Fig.~\ref{fig:PLATO_fov}), that is, we assume 7500 (or 10,000) stars in each of the two fields. Hence, the (2\,yr~+~2\,yr) strategy will contribute 15,000 stars in the conservative and 20,000 stars in the optimistic scenario. In contrast, the (3\,yr~+~1\,yr) strategy will only contribute 7500 (or 10,000) stars to our experiment since we assume that the 1\,yr field will not yield any periodic transits (only mono transits) of Earth-like planets in the HZs around Sun-like stars.

As for our neglect of finding mono transits in PLATO's hypothetical 1\,yr long-duration observation field, {\tt TLS} (like {\tt BLS} and similar phase-folding algorithms) is most sensitive to periodic signals. Although large-scale human inspection of thousands of light curves from the Kepler \citep{2013ApJ...776...10W,2015ApJ...815..127W} and K2 missions \citep{2018RNAAS...2...28L} have revealed about 200 mono transit candidates in total, the corresponding planets are all substantially larger than Earth. There have also been serendipitous mono transit discoveries with dedicated follow-up observations to confirm rare long-period transiting candidates as planets \citep{2018A&A...615L..13G} and progress has been made on the methodological side with the development of new automatized search tools for mono transits \citep{2016AJ....152..206F,2016MNRAS.457.2273O}. But in none of these cases mono transits from Earth-sized planets have been shown to be detectable. 

The geometric transit probability of an exoplanet can be approximated as $P_{\rm geo}~{\sim}~R_{\rm s}/a$. In our case of planets at 1\,AU around solar-type stars we have $R_\odot/{\rm 1\,AU}~{\sim}~0.5$\,\%. We thus expect PLATO to observe (but not necessarily deliver detectable signal of) $15,000~\times~37\,\%~\times~0.5\,\%\,\sim\,28$ transiting planets with $0.5\,R_\oplus~\leq~R_{\rm p}~\leq~1.5\,R_\oplus$ and orbital periods between 340\,d and 760\,d (within the conservative HZ) around F5-K7 dwarf stars in the conservative scenario. In the optimistic scenario, we expect observations of $20,000~\times~88\,\%~\times~0.5\,\%~\sim~88$ such planets. The question we want to answer with our transit injection and retrieval experiment in simulated PLATO data is how many of these transits a state-of-the-art transit detection software would be able to detect.

\begin{figure*}[h]
    \centering
    \includegraphics[width=0.497\linewidth]{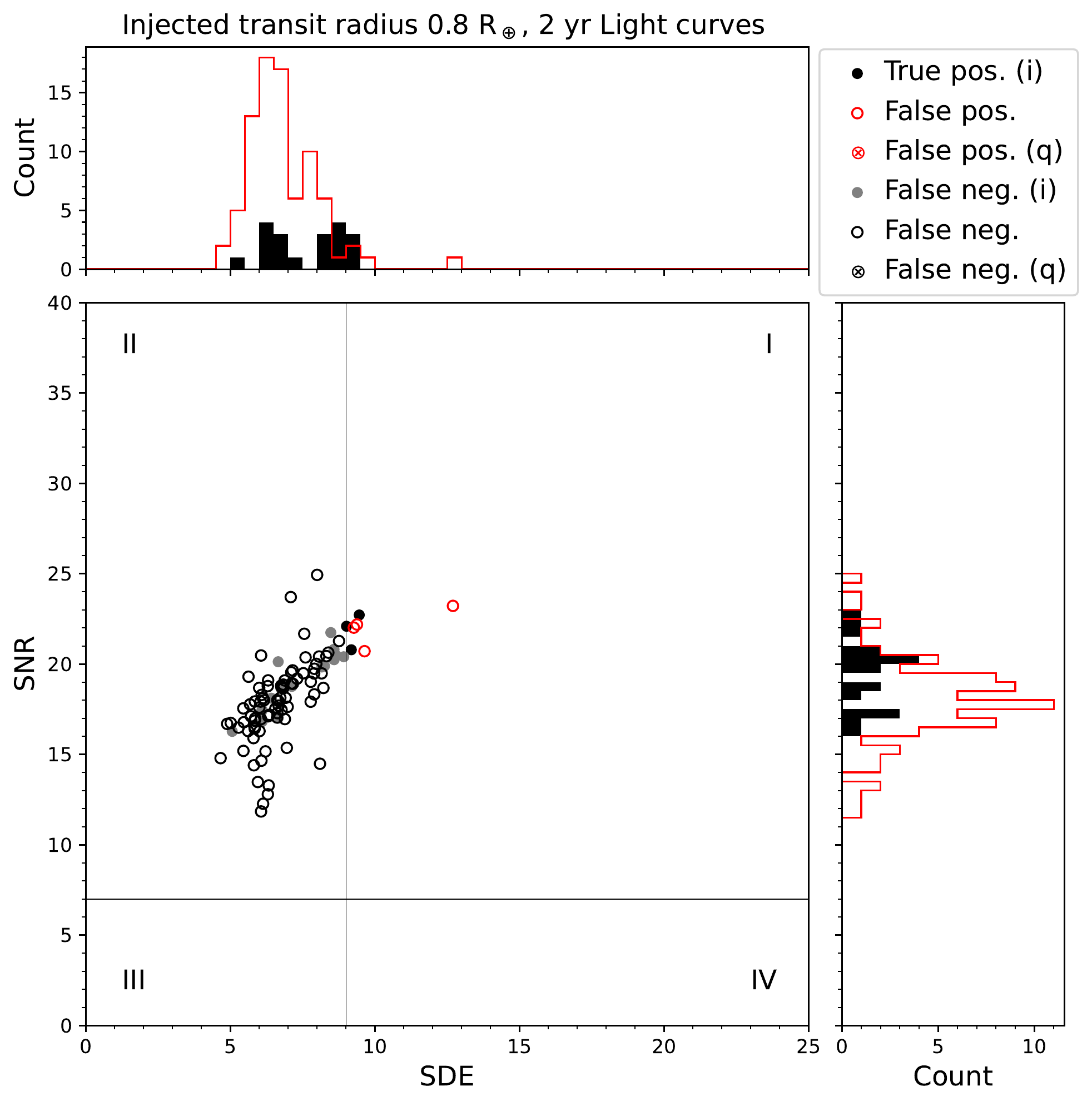}
    \includegraphics[width=0.497\linewidth]{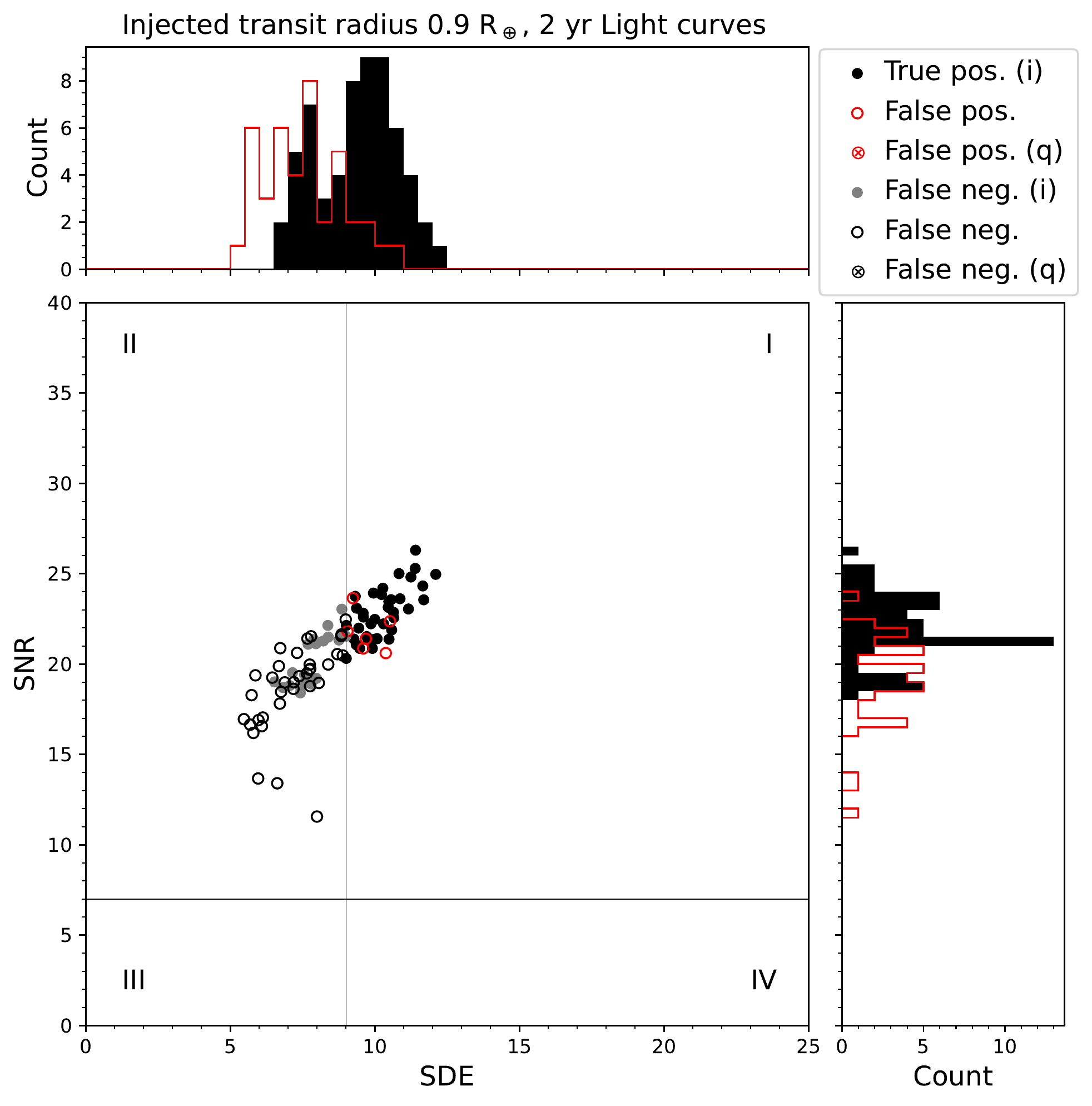}
    \caption{SDE vs. S/N distribution obtained with {\tt TLS} for 100 simulated 2\,yr light curves from 24 PLATO cameras for bright ($m_V=8$) Sun-like stars. (a) Light curves containing two transits of an $R_{\rm p}=0.8\,R_\oplus$ planet.  (b) Light curves containing two transits of an $R_{\rm p}=0.9\,R_\oplus$ planet. Black lines at $\mathrm{SDE}=9$ and $\mathrm{S/N}=7$ define our nominal detection threshold. Black dots in quadrant I are true positives. Red open circles in quadrant I are false positives that are not related to quarterly jumps in the light curves. Red crossed circles in quadrant I are related to false positives caused by quarter jumps (not present in this plot but in similar plots in Appendix~\ref{app:SDE-SNR_injections}). Gray dots outside of quadrant I are recoveries of the injected transit sequence but with $\mathrm{SDE}<9$ or $\mathrm{S/N}<7$ or both, that is, false negatives. Black open circles outside of quadrant I are false negatives not related to the injected signals. Black crossed circles outside of quadrant I are false negatives related to quarter jumps. The ``(i)'' in the legend refers to the injected signal, and ``(q)'' refers to quarter jumps. In the histograms, solid black bars count the number of injected signals recovered in the respective SDE or S/N bins, whereas red line bars count the number of false detections (both positives and negatives).}
    \label{fig:SDE_SNR_comparison_08r_09r}
\end{figure*}

\subsection{PLATO field of view and star count}
\label{sec:fov}

For the purpose of computing PLATO's expected planet yield, we weight the planet detection rates for different camera counts (6, 12, 18, 24) with the relative areas of the camera count coverage on the celestial plane. The combined field of view of PLATO's 24 normal cameras is 2132\,deg$^2$. The provisional pointings of the long-duration observation fields (NPF and SPF)\footnote{The provisional NPF has been designed to overlap with the field of the Kepler primary mission. Though not a primary focus of PLATO, this overlap might allow for the validation of a significant number of Kepler exoplanet candidates and could help to secure exoplanet ephemeris.} as well as the provisional pointings during the step-and-stare phase are shown in Fig.~\ref{fig:PLATO_fov}. Both the NPF and SPF are centered around a Galactic latitude of $30^\circ$. Though not perfectly identical, the stellar densities in the NPF and SPF are extremely similar \citep[Fig.~8 in][]{2022A&A...658A..31N}.
The central area of the PLATO field, which is covered by all 24 normal cameras, is 294\,deg$^2$ ($13.8\%$ of the combined field of view). The area covered by 18 cameras has a size of 171\,deg$^2$ ($8.0\%$), by 12 cameras 796 deg$^2$ ($37.3\%$), and by 6 cameras 871\,deg$^2$ ($40.9\%$) \citep{PLATO_fov_update}.

We also weight the planet detection rates with the relative star count as a function of apparent magnitude in these areas. We examine the Gaia Early Data Release 3 \citep[EDR3;][]{2016A&A...595A...1G,2021A&A...649A...1G} and find a total of 1,247,240 stars with a $G$-band magnitude $m_G~\leq~11$ in the NPF and SPF. We then broke the Gaia EDR3 star counts into the following magnitude bins, assuming similar relative star counts in the PLATO wavelength band:

\vspace{-0.4cm}

\begin{align*}
    m_V &= 8: \hspace{1.18cm} m_G < 8.5 & \Rightarrow \ \ \ \ 106,364\, \mathrm{stars} \ \  \ (8.5\,\%)\\
    m_V &= 9: \hspace{0.32cm} 8.5 \leq m_G < 9.5 & \Rightarrow \ \ \ \  187,884\, \mathrm{stars} \ (15.1\,\%)\\
    m_V &= 10: \hspace{0.14cm} 9.5 \leq m_G < 10.5 & \Rightarrow \ \ \ \ 484,765\, \mathrm{stars} \ (38.9\,\%)\\
    m_V &= 11: 10.5 \leq m_G \leq 11 & \Rightarrow \ \ \ \ 468,228\, \mathrm{stars} \ (37.5\,\%).
\end{align*}

\noindent
The purpose of this binning was to obtain the weighting factors for the TPRs from our injection-retrieval tests that include stars of different magnitude. As for the spectral types, the distribution of stellar spectral types in the NPF and SPF was unknown at the time of our writing, and the all-sky PLATO input catalog has only been published very recently \citep{2021A&A...653A..98M}. It would not have been helpful to take spectral type counts from Gaia because additional conditions are applied to the selection of the P1 sample stars in addition to magnitude cuts. Hence, we chose a Sun-like star for reference. This choice follows the procedure of using Sun-like benchmark targets in the PLATO Definition Study Report (p. 111 therein). This approximation certainly affects the expected planet yields significantly, the extent of which could be examined in future studies.

\subsection{Transit injection-and-retrieval experiment}

We create two sets of light curves with {\tt PSLS}. In the pollinated set, each light curve contains a transit sequence of one planet. In the control set, no light curve contains a transit. From the pollinated sample we determine the TPR, FPR, and false negative rate (FNR). The control set is used to study the FPR and the true negative rate (TNR).

For the pollinated set, we created 2\,yr and 3\,yr light curves with transits for five planet radii $R_{\rm p}~\in~\{0.8, 0.9, 1.0, 1.2, 1.4\}~\times~R_\oplus$, four reference magnitudes $m_V\,{\in}\,\{8,9,10,11\}$, and assuming 24 cameras, each combination of which is represented by 100 light curves. This gives a total of $2~{\times}~5~{\times}~4~{\times}~100~=~4000$ light curves.

For Earth-sized planets only, we also simulated 2\,yr and 3\,yr light curves for all combinations of stellar apparent magnitudes $m_V\,{\in}\,\{8,9,10,11\}$ and camera counts $n_{\rm cams}\,{\in}\,\{6,12,18,24\}$, giving another $2~{\times}~4~{\times}~4~{\times}~100~=~3200$ light curves. For the control set, which does not contain any transits, we compute 2\,yr and 3\,yr light curves for $m_V\,{\in}\,\{8,9,10,11\}$ and $n_{\rm cams}\,{\in}\,\{6,12,18,24\}$, leading again to $2~{\times}~4~{\times}~4~{\times}~100~=~3200$ light curves. All things combined, we generated $10,400$ light curves with {\tt PSLS}.

For all simulated transits, we set the transit impact parameter to zero, which means that we assumed that all transits occur across the entire diameter. We also used $t_0 = 91$\,d and $P=364$\,d throughout our simulations; the latter choice was made to prevent a systematic effect at the Earth's orbital period of 365.25\,d that was apparent in some {\tt PSLS} light curves.

We execute our computations on a ($4{\times}12$)-core AMD Opteron(tm) 6174 processor (2200\,MHz) on a GNU/Linux operating system with a 64-bit architecture. Computation of a 2\,yr (3\,yr) light curve with {\tt PSLS} takes about 5\,min (9\,min), depending on the configuration. The detrending of a 2\,yr (3\,yr) light curve with W{\={o}}tan typically takes 15\,min (25\,min). The transit search in each 2\,yr (3\,yr) light curve with {\tt TLS} includes periods between 50.0\,d and 364.75\,d (547.25\,d), the upper boundary chosen to embrace the most long-period planets that exhibit at least two transits, and typically takes 4.5\,h (10\,h). These numbers were measured during parallel processing of typically around ten jobs, giving a total CPU time of about 270\,d.

\begin{figure*}[h]
    \centering
    \includegraphics[width=0.497\linewidth]{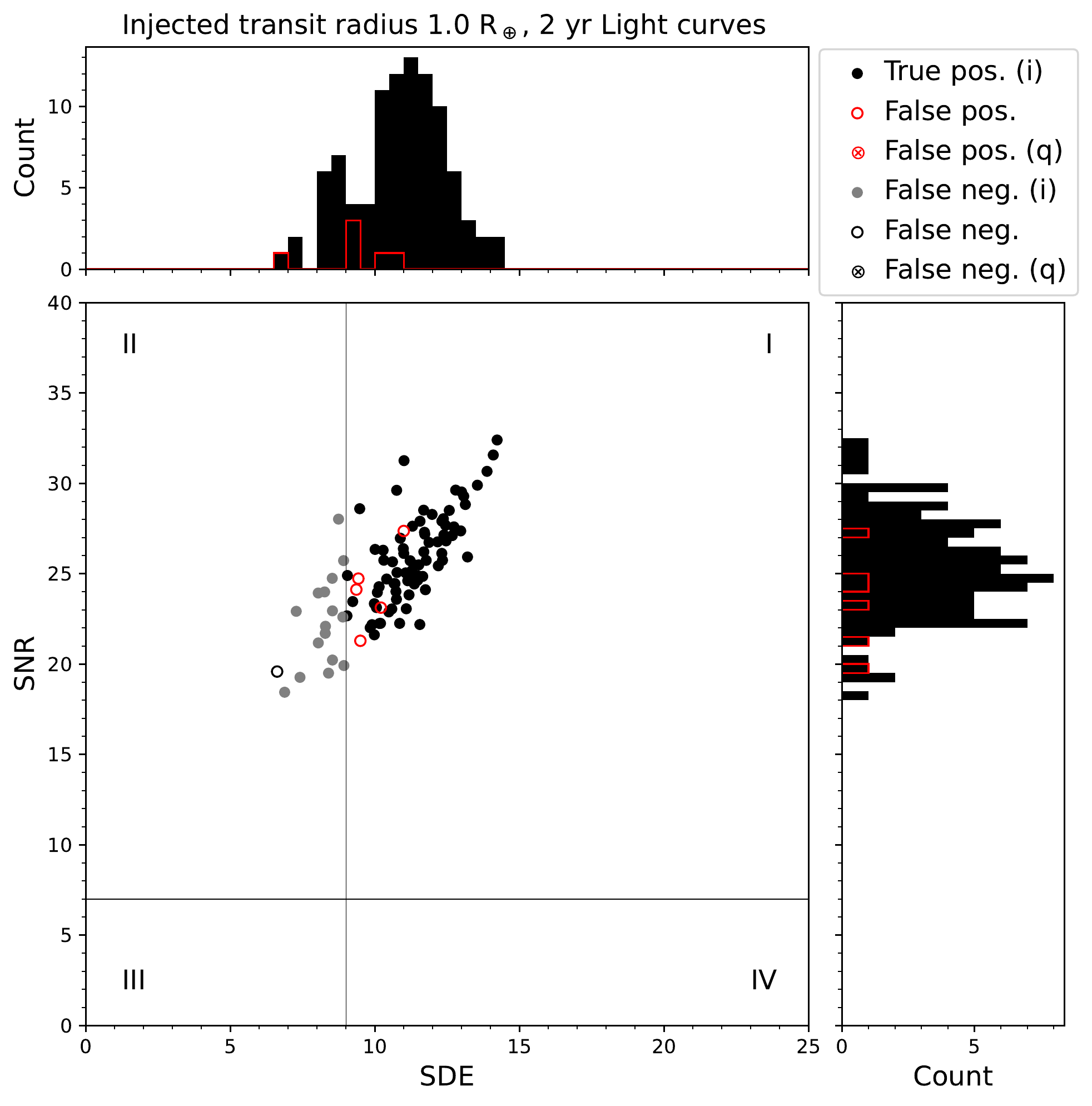}
    \includegraphics[width=0.497\linewidth]{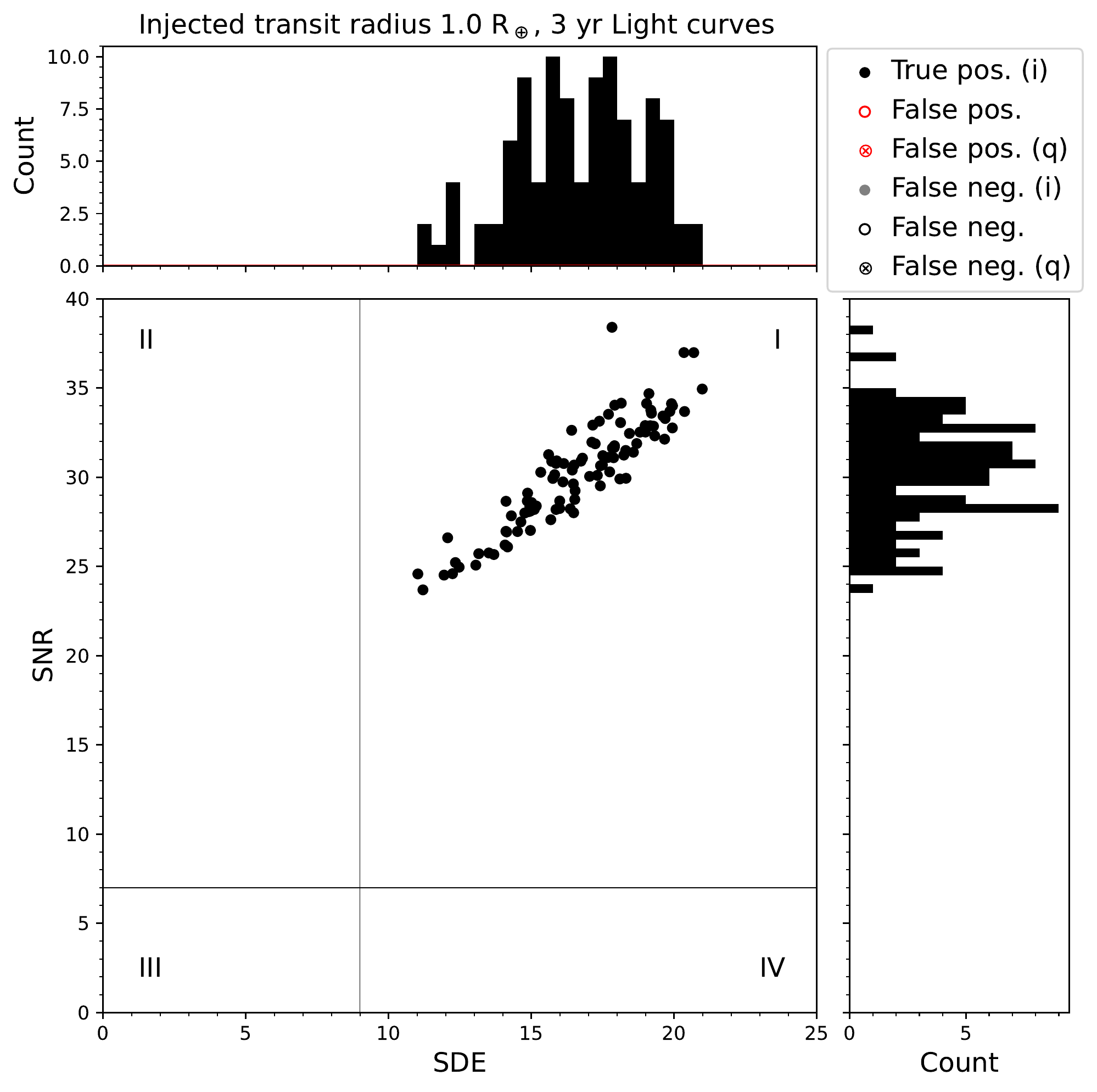}
    \caption{Similar to Fig.~\ref{fig:SDE_SNR_comparison_08r_09r} but now for Earth-sized planets around bright ($m_V=8$) Sun-like stars and 24 PLATO cameras. Panel (a) is based on two transits in 100 light curves of 2\,yr length, and panel (b) shows the results for three transits in 100 light curves that are 3\,yr long.}
    \label{fig:SDE_SNR_10r_2y_3y}
\end{figure*}

\begin{figure*}
    \centering
    \includegraphics[width=\linewidth]{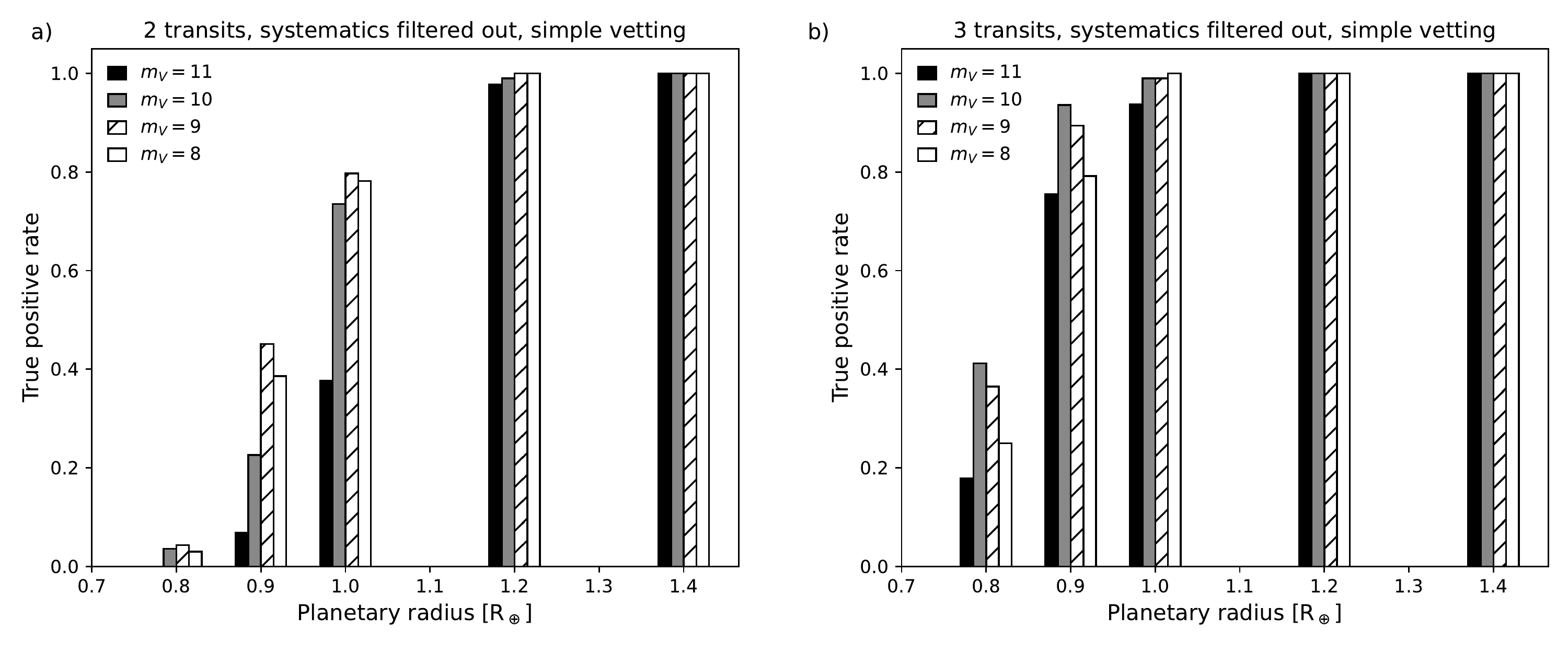}
    \caption{True positive rate of our injection-and-retrieval experiment in simulated light curves from 24 PLATO cameras as a function of injected transiting planet radius. All stars are assumed to be Sun-like. A total of 100 light curves were generated, detrended, and searched for transits for each histogram bar. Detections are defined as any recovery of an injected signal within ${\pm}1$\,d of the injected $P$ or $T_0$ values and if the injected signal was detected by {\tt TLS} as the strongest signal with ${\rm SDE}~{\geq}~9$ and ${\rm S/N}~{\geq}~7$. Results for four apparent stellar magnitudes are shown in different histogram shadings (see legend). In panel (a) simulated light curves have a length of 2\,yr. In panel (b) light curves have a length of 3\,yr.}
    \label{fig:TPR_Radius}
\end{figure*}

\section{Results}

\subsection{Detection rates}
\label{sec:rates}

In Fig.~\ref{fig:SDE_SNR_comparison_08r_09r} we show the SDE versus S/N distribution obtained with {\tt TLS} for 100 light curves that cover 2\,yr from 24 PLATO cameras. Each light curve featured one transiting planet in a 364\,d orbit around an $m_V~=~8$ Sun-like star. Panel (a) refers to planets with a radius of $0.8\,R_\oplus$, panel (b) refers to $R_{\rm p}=0.9\,R_\oplus$. Although {\tt TLS} has the ability to search for successively weaker signals in a given light curve, we stop our search after the first detection of the strongest signal. Quadrant I, where $\mathrm{SDE}~{\geq}~9$ and $\mathrm{S/N}~{\geq}~7$, contains the true positives (black filled dots), that is, the retrievals of the injected transits. We define two types of false positives. Red open circles in quadrant I represent false positives not related to quarter jumps and red crossed circles illustrate false positives caused by quarter jumps. Moreover, we define three types of false negatives. One such type is a detection of the injected signal with $\mathrm{SDE}<9$ or $\mathrm{S/N}<7$ or both (gray filled dots). False negatives can also be unrelated to the injected signal and instead be caused by {\tt TLS}'s detection of systematic or astrophysical variability below our nominal SDE-S/N detection threshold. If such a false negative does not refer to a detection of a quarter jump, we plot it with a black open circle. If a false negative is caused by a quarter jump outside of quadrant I, we illustrate it with a black crossed circle. The histograms along the ordinate and abscissa illustrate the SDE and S/N distributions of the injected signal recoveries (black filled bars) and false detections (red line bars), respectively.

\begin{figure*}
    \centering
    \includegraphics[width=\linewidth]{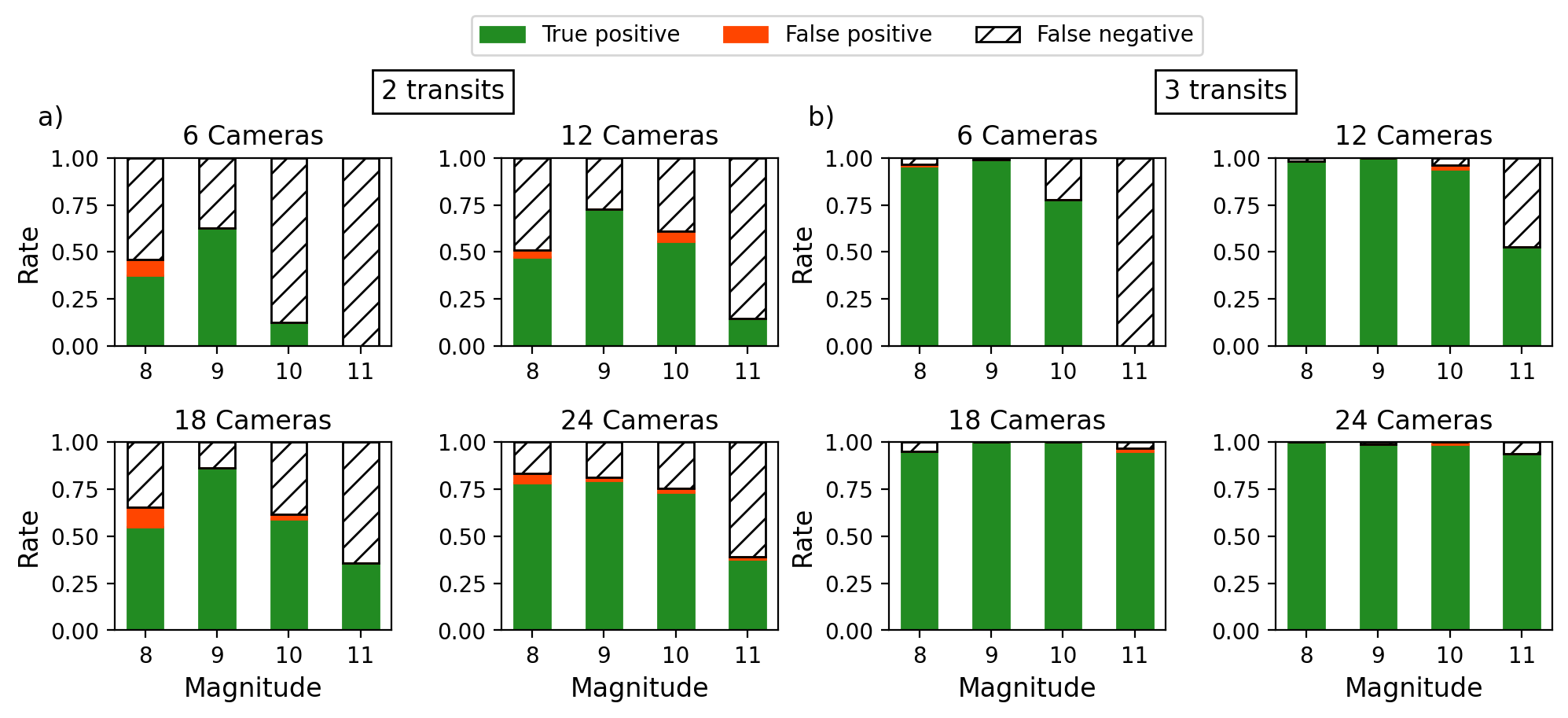}
    \caption{Summary of our transit injection-and-retrieval experiment with PLATO-like light curves of Sun-like stars with an Earth-sized transiting planet in a one-year orbit. As true positives, we count {\tt TLS} detections that fulfill the same criteria as those in Fig.~\ref{fig:TPR_Radius}. (a) Simulated light curves from {\tt PSLS} with a duration of 2\,yr and containing two transits. (b) Simulated light curves from {\tt PSLS} with a duration of 3\,yr and containing three transits.}
    \label{fig:TPR_Mag}
\end{figure*}

In comparing Figs.~\ref{fig:SDE_SNR_comparison_08r_09r}(a) and (b), we recognize a substantial shift of the SDE versus S/N distribution toward higher values, respectively. As a consequence, the TPR increases from 3\,\% in panel (a) to 39\,\% in panel (b). The FPR (red symbols in quadrant I) is 4\,\% in (a) and 6\,\% in (b). As a result, the detection reliability, which we define as $R={\rm TPR/(TPR + FPR)}$, is $3/7 \sim 43\,\%$ in panel (a) and $R = 39/45 \sim 87\,\%$ in panel (b).

In Fig.~\ref{fig:SDE_SNR_10r_2y_3y} we move on to Earth-sized planets ($R_{\rm p}=R_\oplus$) in one-year orbits around bright ($m_V=8$) Sun-like stars.\footnote{The SDE vs. S/N distributions obtained for injection and retrieval experiments with $R_{\rm p}=R_\oplus$ for both 2\,yr and 3\,yr light curves of Sun-like stars with $m_V~\in~\{9,10,11\}$, all of which assumed 24 cameras, are available in Appendix~\ref{app:SDE-SNR_injections}.} Panel (a) represents 100 simulated light curves obtained with 24 PLATO cameras over 2\,yr (similar to Fig.~\ref{fig:SDE_SNR_comparison_08r_09r}), whereas panel (b) shows our results for 3\,yr light curves. In our analysis of the 2\,yr light curves, each of which contains two transits, we obtain TPR~=~78\,\%, FNR~=~17\,\%, FPR~=~5\,\%, and $R=94\,\%$. For comparison, in Fig.~\ref{fig:SDE_SNR_10r_2y_3y}(b), where each light curve contained three transits, we find TPR~=~100\,\% and $R=100\,\%$.

As a major result of this study, we find that all of the injected transits of Earth-sized planets in Earth-like orbits around $m_V=8$ Sun-like stars are recovered when three transits are available. In fact, the increase in both SDE and S/N is significantly more pronounced when moving from two to three transits compared to increasing the planetary radius from $0.8\,R_\oplus$ to $1\,R_\oplus$. Moreover, our measurements of the detection reliability suggests that $R \sim 100\,\%$ for Earth-sized planets and larger.

\begin{figure*}
    \centering
    \includegraphics[width=\linewidth]{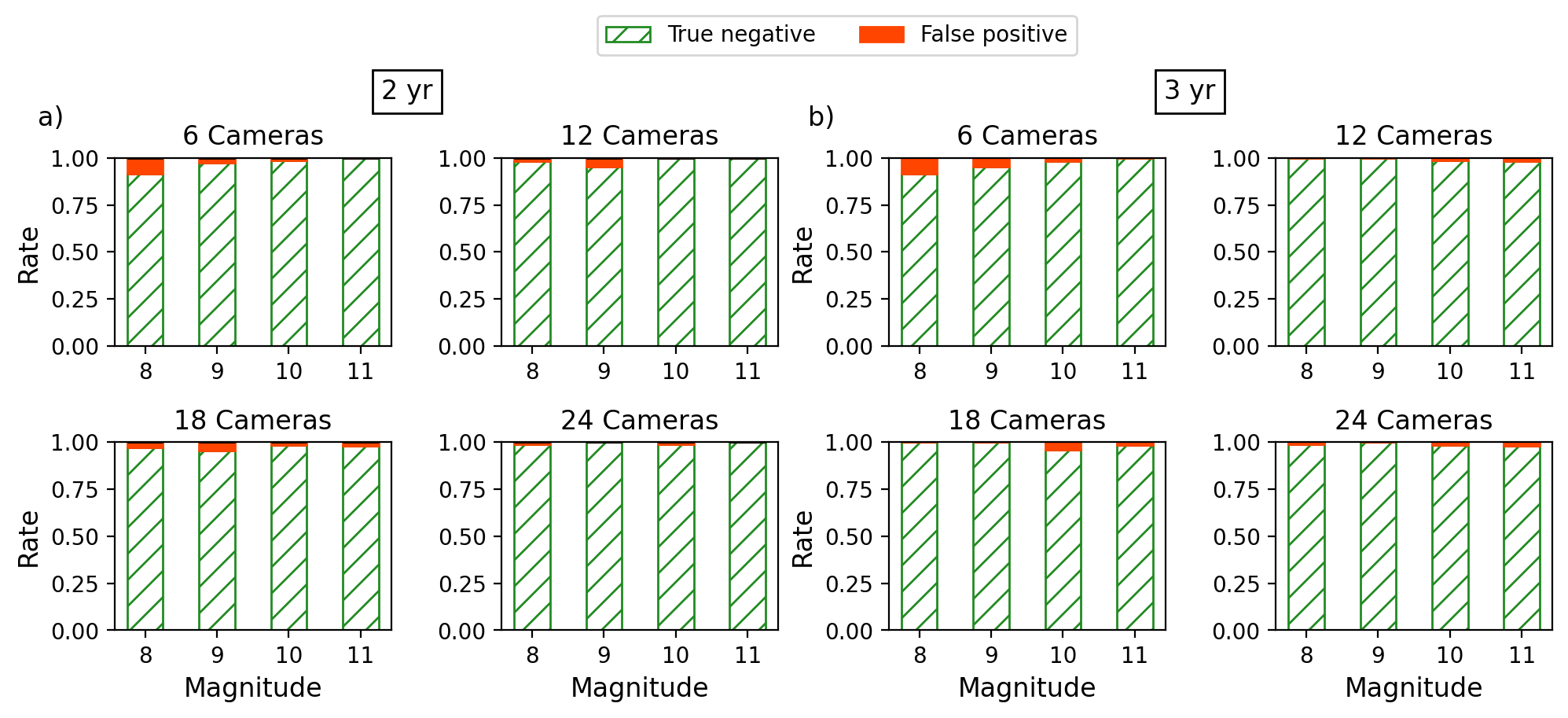}
    \caption{Summary of our transit injection-and-retrieval experiment with PLATO-like light curves of Sun-like stars without any transiting planet. (a) Simulated light curves from {\tt PSLS} with a duration of 2\,yr. (b) Simulated light curves from {\tt PSLS} with a duration of 3\,yr.}
    \label{fig:TNR_Mag}
\end{figure*}

In Fig.~\ref{fig:TPR_Radius} and the following plots in this paper, we neglect any false positives detected with periods near the PLATO quarters, that is, with $P=90~({\pm}1)$~d and full-integer multiples of that. {\tt W{\={o}}tan} does a decent job at removing quarterly jumps (see Fig.~\ref{fig:method_overview}), but real PLATO data will be cleaned from gaps and jumps prior to any transit search, which is why this sort of false positives will be drastically reduced in the analysis of real PLATO data. The histograms in Fig.~\ref{fig:TPR_Radius} illustrate the TPRs as a function of the injected planet radius ($R_{\rm p}~\in~\{0.8, 0.9, 1.0, 1.2, 1.4\}~\times~R_\oplus$). For each planetary radius, we show four bars, which refer to $m_V~\in~\{11,10,9,8\}$ from left to right, respectively. All light curves were simulated with 24 PLATO cameras. Panel (a) refers to 2\,yr light curves with two transits, and panel (b) refers to 3\,yr light curves with three transits.

As a general trend, and as expected, we find that the TPRs increase with increasing planetary radius in both panel (a) and (b). In panel (a), the TPRs of the 2\,yr light curves are on the percent level for the smallest planets that we investigated ($R_{\rm p}~=~0.8\,R_\oplus$), irrespective of the stellar apparent magnitude. For $R_{\rm p}=0.9\,R_\oplus$, the TPRs in (a) increase from $9\,$\% for $m_V=11$ to $20\,$\% for $m_V=10$ and $50\,$\% for $m_V=9$. Interestingly, the brightest stars ($m_V=8$) do not yield the largest TPRs ($39\,$\%), which we attribute to the associated saturation of the CCD pixels. This loss of true positives due to saturation is only present for sub-Earth-sized planets in panel (a). For Earth-sized planets, the TPR is about 28\,\% for $m_V=11$ and near 80\,\% for $m_V~\in~\{10,9,8\}$. For planets as large as or larger than $1.2\,R_\oplus$ the TPRs are near 100\,\%.

For comparison, in panel (b) the TPRs of the 3\,yr light curves are between about 20\,\% and 45\,\% for $R_{\rm p}=0.8\,R_\oplus$, which represents a substantial boost from the near zero TPRs derived for these small planets in the 2\,yr light curves in panel (a). Moreover, the TPRs in panel (b) reach 80\,\% for $R_{\rm p}=0.9\,R_\oplus$ and between 91\,\% and 100\,\% for $R_{\rm p}=\,R_\oplus$, details depending on the apparent stellar magnitude. For planets as large or larger than $1.2\,R_\oplus$ the TPRs are 100\,\% for all apparent stellar magnitudes.

In Fig.~\ref{fig:TPR_Mag} we focus on Earth-sized planets in one-year orbits around bright ($m_V=8$) Sun-like stars, showing the TPRs, the FPRs, and the FNRs. Panel (a) refers to 2 yr light curves with two transits, and panel (b) refers to 3 yr light curves with three transits. In each panel, four histograms illustrate the respective rates for 6, 12, 18, and 24 PLATO cameras (see histogram titles).

In general, the highest TPRs and lowest FNRs in Fig.~\ref{fig:TPR_Mag} are obtained for the largest camera counts and lowest apparent stellar magnitudes. As a general result we find that TPRs are almost always $<75\,\%$ based on two transits in the 2\,yr light curves in panel (a), even for 24 cameras and $m_V~\in~\{8,9,10\}$. For comparison, for three transits in the 3\,yr light curves in panel (b), ${\rm TPR}~{\approx}~100\,\%$ in almost all the cases that we investigated, except for {\tt PSLS} light curves simulated with 6 cameras of moderately dim stars with $m_V~\in~\{10,11\}$ and for light curves from 12 cameras and $m_V=11$. 
For systems with three transits, TPRs are near 100\,\% for $m_V\,\in\,\{8,9\}$ and about 75\,\% for $m_V\,=\,10$ even for as few as six PLATO cameras. As a consequence, an increase in the camera count is only beneficial to the TPRs of Earth-sized planets with three transits around stars with $m_V\,=\,11$ (and likely for even dimmer stars).

Figure~\ref{fig:TPR_Mag} also demonstrates that the TPRs increase substantially for targets with low camera counts ($n_{\rm cams}\,{\in}\,\{6,12\}$) when adding a third transit. This is particularly important given the fact that most of the PLATO field of view is covered with 6 cameras ($40.9\%$) or 12 cameras ($37.3\%$) (see Fig.~\ref{fig:PLATO_fov} and Sect.~\ref{sec:fov}). As a consequence, these areas have the greatest impact on the expected planet yields. Adding observations of a third transit is extremely beneficial to the planet yields in these regions of the PLATO field of view that are covered by 6 or 12 cameras.

Figure~\ref{fig:TNR_Mag} illustrates the TNRs and FPRs obtained with {\tt TLS} from {\tt PSLS} light curves without any injected signal. Panel (a) refers to 2\,yr light curves, panel (b) refers to 3\,yr light curves. As a general result, FPRs for the 2\,yr light curves in (a) are on the percent level at most and we do not observe any significant correlation between the FPRs and the camera count or the FPRs and the stellar apparent magnitude. For the 3\,yr light curves in (b), our interpretation is similar except for a dramatic increase in the FPRs to 100\,\% for $n_{\rm cams}=6$ and $m_V=8$. In this particular configuration, saturation is combined with a particularly weak transit signal and so {\tt TLS} is more sensitive to the quarter jumps than to the transit.

\subsection{Earth-like planet yield}

Next, we predict PLATO's yield of Earth-like planets in the HZ around Sun-like stars. To this purpose, we first interpolate the TPRs derived from light curves with 24 PLATO cameras to cover the entire range of planetary radii we are interested in. For the 2\,yr light curves, we use TPR~=~0 for $R_{\rm p}\,\leq\,0.7\,R_\oplus$ and for the 3\,yr light curves, we use TPR~=~0 for $R_{\rm p}\,\leq\,0.6\,R_\oplus$ for all magnitudes, respectively (illustrated in Fig.~\ref{fig:TPR_Radius}). We then use our results for the TPRs of Earth-sized planets (Fig.~\ref{fig:TPR_Mag}) to extrapolate the dependence of the TPRs on the camera counts across all magnitudes and planetary radii. This results in a distribution of the TPRs as a function of $R_{\rm p}$, $m_V$, and $n_{\rm cams}$.

As an aside, we find that under the most unfavorable conditions of $n_{\rm cams}=6$ and $m_V=11$, TPRs near 100\,\% are obtained for planets as small as $1.5\,R_\oplus$ for two transits and as small as $1.3\,R_\oplus$ for three transits.

We then multiply the resulting TPRs with our estimates of observable transiting Earth-like planets from Sect.~\ref{sec:fov}. We consider both the (2\,yr~+~2\,yr) and the (3\,yr~+~1\,yr) strategies for PLATO's long-duration observation phase (see Sect.~\ref{sec:occurrence}). Although {\tt TLS} can detect mono transits, its sensitivity is strongly diminished compared to periodic signals. Hence, we neglect mono transits in our analysis and assume that no transits will be detected during the $1\,\mathrm{yr}$ observation phase (see Sect.~\ref{sec:occurrence}).

Our predictions of PLATO's Earth-like planet yield are shown in Tables~\ref{tab:planet_yield_mag} and ~\ref{tab:planet_yield_cams}. In Table~\ref{tab:planet_yield_mag}, planet counts are itemized as per the apparent stellar magnitude bins in which we predict them (see Sect.~\ref{sec:fov}), whereas in Table~\ref{tab:planet_yield_cams} planet counts are shown as a function of the PLATO camera count. The conservative and optimistic scenarios refer to different assumptions of the star count and HZ exoplanet occurrence rate as detailed in Sect.~\ref{sec:occurrence}. Although we are fully aware that only a full-integer number of exoplanets can be found, we chose to present our predictions including one decimal place given the low number counts.

\begin{table}[h]
\caption{Estimated yield of planets with $0.5\,R_\oplus~\leq~R_{\rm p}~\leq~1.5\,R_\oplus$ in the HZs around Sun-like stars from PLATO's P1 sample.}
\def\arraystretch{1.1}
\label{tab:planet_yield_mag}
\centering
    \begin{tabular}{c|c|c|c}
    \hline
    Observing & $m_V$ & Conservative & Optimistic \\
    Strategy       &             & scenario & scenario \\
    \hline
        2 yr + 2 yr & 8 & 1.2 & 3.4 \\
                    & 9 & 2.2 & 6.9 \\
                    & 10 & 4.4 &  14.0 \\
                    & 11 & 3.0 & 9.6 \\\hline
            & total & {\bf 10.7} & {\bf 33.8} \\\hline\hline
        3 yr + 1 yr & 8 & 0.8 & 2.4 \\
                    & 9 & 1.4 & 4.3 \\
                    & 10 & 3.4 &  10.8 \\
                    & 11 & 2.5 & 7.9 \\\hline
            & total & {\bf 8.0} & {\bf 25.4} \\
    \hline\hline
    \end{tabular}\\
\tablefoot{The two different observing strategies assume that the two long-duration observation phase fields are monitored for either (2\,yr~+~2\,yr) or (3\,yr~+~1\,yr), respectively. The conservative (optimistic) scenario assumes 15,000 (20,000) Sun-like stars in the P1 sample and an occurrence rate of 0.37 (0.88) Earth-sized planets in the HZ per star. 
Planet yields are sorted in bins of apparent stellar magnitudes (see Sect.~\ref{sec:fov}). Numbers are rounded to one decimal place, which leads to an offset of 0.1 between some of the column sums and the respective total counts. Total counts are exact.}
\end{table}

The key insight to be gained from Tables~\ref{tab:planet_yield_mag} and ~\ref{tab:planet_yield_cams} is that the (2\,yr~+~2\,yr) observing strategy produces significantly higher planet yields than the (3\,yr~+~1\,yr) observing strategy in any scenario. This interpretation is supported by the total planet yield counts in both the conservative and optimistic scenarios. The total count is 10.7 for the (2\,yr~+~2\,yr) strategy compared to 8.0 for the (3\,yr~+~1\,yr) strategy in the conservative scenario. In the optimistic scenario, the (2\,yr~+~2\,yr) strategy produces a predicted yield of 33.8, whereas the (3\,yr~+~1\,yr) strategy yields 25.4 planets. Details of the conservative versus optimistic scenarios aside, the yield of the (2\,yr~+~2\,yr) strategy is 133\,\% of the (3\,yr~+~1\,yr) strategy.

In addition to these actual discoveries of small planets in the HZs around Sun-like stars, our results suggest a detection reliability near 100\,\% for Earth-sized and larger planets (see Sect.~\ref{sec:rates}). Hence, we do not expect a significant amount of statistical false detections, that is, false positives caused by systematic, instrumental, or statistical effects for super-Earth-sized planets. In fact, \citet{Hippke_Heller_2019_tls} showed that an SDE threshold of 9 for {\tt TLS} produces one statistical false positive in about 10,000 light curves with normally distribute noise with an amplitude of 30\,ppm. Consequently, the P1 sample with its 15,000 - 20,000 stars will yield 1-2 statistical false positives. That said, there will be about as many false detections of transiting planets smaller than Earth as there will be genuine sub-Earth-sized planets. And on top of that, there will also be false positives caused by astrophysical effects such as blended eclipsing binaries.

\section{Discussion}

\subsection{Effects of observing strategy, scenario, and detection thresholds}

We find that the choice of the observing strategy is not as impactful as the realization of the conservative versus the optimistic scenario. The realization of the conservative or the optimistic scenario can only be affected to a limited extent, for example through the choice of the PLATO observing fields. Although neither the dilution of exoplanet transits due to stellar blending nor the occurrence of astrophysical false positives (e.g., from blended eclipsing stellar binaries) have been taken into account in our simulation, this issue has been taken care of by the PLATO consortium by optimizing the trade-off between a high number of priority targets and a low number of false-alarm detections due to crowding \citep{2014ExA....38..249R,2022A&A...658A..31N}. The exoplanet occurrence rate, however, which also goes into our definition of the conservative versus optimistic scenarios, is an observational property and needs to be taken as is. 

\begin{table}[t]
\caption{Estimated planet yields as in Table~\ref{tab:planet_yield_mag} but yields are sorted in terms of the number of cameras used for the simulated light curves.}
\def\arraystretch{1.1}
\label{tab:planet_yield_cams}
\centering
    \begin{tabular}{c|c|c|c}
    \hline
    Observing & cams & Conservative & Optimistic \\
    Strategy       &             & scenario & scenario \\
    \hline
        2 yr + 2 yr &  6 & 3.5 & 11.2 \\
                    & 12 & 4.1 & 13.1 \\
                    & 18 & 1.0 &  3.3 \\
                    & 24 & 2.0 & 6.3 \\\hline
            & total & {\bf 10.7} & {\bf 33.8} \\\hline\hline
        3 yr + 1 yr &  6 & 3.0 & 9.4 \\
                    & 12 & 3.1 & 9.9 \\
                    & 18 & 0.7 &  2.3 \\
                    & 24 & 1.3 & 4.0 \\\hline
            & total & {\bf 8.0} & {\bf 25.4} \\
    \hline\hline
    \end{tabular}
\end{table}

In our injection-and-retrieval experiment, we used SDE~=~9 and S/N~=~7 as our nominal detection thresholds. The SDE versus S/N distributions in Figs.~\ref{fig:SDE_SNR_10r_2y_3y}(b) and Figs.~\ref{fig:SDE_SNR_10r_2y_3y_mag9}(b)-\ref{fig:SDE_SNR_10r_2y_3y_mag11}(b) as well as the summary plot in Fig.~\ref{fig:TPR_Radius}(b) show that these thresholds are sufficient to detect Earth-sized (and larger) planets using three transits around bright and moderately bright ($m_V\,{\leq}\,11$) Sun-like stars with ${\rm TPR}\,>\,90\,\%$. Moreover, Figs.~\ref{fig:SDE_SNR_no_2y_3y_mag8}-\ref{fig:SDE_SNR_no_2y_3y_mag11} illustrate that most false signals achieve ${\rm SDE}\,<\,9$ in the first transit search with {\tt TLS}, although the S/N is often substantially above 10. The only type of false alarm signal with ${\rm SDE}\,>\,9$ that we observed in our simulations is quarter jumps, but these can be identified and dismissed. There will be other sources of false positives for PLATO, but their quantification is beyond the scope of this study. As a tribute to a rigid set of SDE and S/N thresholds, sub-Earth-sized planets are hard to be discriminated from false alarms, as becomes clear in a comparison of the SDE versus S/N distribution of the injected signals in Fig.~\ref{fig:SDE_SNR_comparison_08r_09r} with the SDE versus S/N distribution of the control sample in Figs.~\ref{fig:SDE_SNR_no_2y_3y_mag8}-\ref{fig:SDE_SNR_no_2y_3y_mag11}. The same tribute is paid for Earth-sized planets with only two transits (see Fig.~\ref{fig:SDE_SNR_10r_2y_3y}(a) and Figs.~\ref{fig:SDE_SNR_10r_2y_3y_mag9}(a)-\ref{fig:SDE_SNR_10r_2y_3y_mag11}(a)). Machine learning methods like self-organizing maps \citep{2017MNRAS.465.2634A}, random forest \citep{2018MNRAS.478.4225A}, or convolutional neural networks \citep{2018MNRAS.474..478P,2020A&A...633A..53O,2021MNRAS.502.2845R} might be helpful in the separation of genuine exoplanet transit signals from false alarms, but for now their advantage over smart-force search algorithms like {\tt TLS} has not been conclusively demonstrated.

\subsection{Planet yields}

Our focus on the strongest transit-like signal (true or false) and our omission of an iterative transit search down to the detection threshold means that we underestimate {\tt TLS}'s capabilities to find shallow transits in PLATO data. In fact, {\tt TLS} can automatically search for successively weaker signals \citep{Hippke_Heller_2019_tls} and there are several ways to interpret an iterative search in terms of true and FPRs. Though this is beyond the scope of this study iterative transit searches will certainly be an important topic for PLATO.

The TPRs for Earth-sized planets transiting $m_V\,=\,8$ Sun-like stars are smaller than for more moderately bright stars with $m_V\,=\,9$ in Figs.~\ref{fig:TPR_Radius} and \ref{fig:TPR_Mag}(a). We attribute this effect to saturation, which results in higher-than-realistic noise levels for the brightest stars. The resulting {\tt PSLS} light curves are thus not representative of real PLATO light curves. That said, these stars are also the least abundant to be observed with PLATO (see Sect.~\ref{sec:fov}). As a consequence, we expect the effect on our expected planets yields to be minor and $\lesssim\,1$ in terms of number counts for all scenarios. For details of the conversion between Johnson's $V$-band magnitude ($m_V$) and PLATO's $P$ magnitude used in {\tt PSLS} (see \citealt{2019A&A...627A..71M}).

A direct comparison of our predicted planet yields in the HZs around Sun-like stars with those presented in the PLATO Definition Study Report is complex due to several reasons. First, this report used analytical estimates of the expected number of planets with ${\rm S/N}\,>\,10$ to predict PLATO's planet yield. For comparison, we used simulated PLATO light curves and a transit injection-and-retrieval experiment. Second, we focused on the P1 stellar sample and chose to represent its 15,000 - 20,000 F5-K7 stars with Sun-like stars, including astrophysical variability. Instead, the PLATO Definition Study Report included $m_V~{\leq}~11$ stars from both the P1 and P5 sample, the latter of which will comprise ${\geq}\,245,000$ F5-K7 dwarf and subgiant stars (assuming two long-duration observation phase field pointings) with $m_V~{\leq}~13$ and a cadence of 600\,s in most cases. Third, the estimate of 6 to 280 small planets in the HZs around $m_V~\leq~11$ stars given in the PLATO Definition Study Report included all planets smaller than $2\,R_\oplus$. By contrast, we derive exoplanet yields for $0.5\,R_\oplus\,\leq\,R_{\rm p}\,\leq\,1.5\,R_\oplus$. Fourth, given the large observational uncertainties at the time, the PLATO Definition Study Report necessarily included a large range of the possible occurrence rates of small planets in the HZ around Sun-like stars of between 2\,\% and 100\,\%. For comparison, our planet yield predictions are based on updated occurrence rates estimates \citep{2021AJ....161...36B}, which define our conservative scenario with 37\,\% and our optimistic scenario with 88\,\%.

Our yield estimates for planets with $0.5\,R_\oplus\,\leq\,R_{\rm p}\,\leq\,1.5\,R_\oplus$ range between 11 in the conservative scenario of the (2\,yr~+~2\,yr) observing strategy (or 8 for the 3\,yr~+~1\,yr observing strategy) and 34 in the optimistic scenario of the (2\,yr~+~2\,yr) observing strategy (or 25 for the 3\,yr~+~1\,yr observing strategy) (see Tables~\ref{tab:planet_yield_mag} and \ref{tab:planet_yield_cams}). With all the caveats of a direct comparison in mind, our range of the predicted yield of small planets in the HZ is much tighter than the previous estimates from the PLATO Definition Study Report and tends to fall in the lower regime of the previous planet yield estimates.

\subsection{Methodological limitations and improvements}
\label{sec:limitations}

Our results demonstrate that the {\tt W{\={o}}tan} detrending software efficiently removes stellar and systematic variability while preserving transit signatures. That said, in some cases we find that {\tt W{\={o}}tan} does not effectively remove quarter jumps from {\tt PSLS} light curves. {\tt W{\={o}}tan} has been designed to detrend jumps by stitching the data prior and after gaps in the light curve, a functionality that can be fine-tuned using the {\tt break\_tolerance} parameter. Real PLATO data will indeed have gaps of at least several hours required for satellite repositioning, which can be stitched with {\tt W{\={o}}tan}. But {\tt PSLS} does not include temporal gaps at quarter jumps for now. This results in occasional false positive detections with {\tt TLS} at these quarter jumps, in particular for stars with $m_V\,{\geq}\,10$ (see Figs.~\ref{fig:SDE_SNR_10r_2y_3y_mag11}, \ref{fig:SDE_SNR_no_2y_3y_mag10}, and \ref{fig:SDE_SNR_no_2y_3y_mag11}).

In the final data products of the actual PLATO mission, quarter jumps will be subjected to a dedicated light curve stitching. As a consequence, this type of false positives will not, or very rarely, be an issue. As explained in Sect.~\ref{sec:rates}, this is why we neglect quarterly false positives in our summary plots (Figs.~\ref{fig:TPR_Radius}-\ref{fig:TNR_Mag}). Nevertheless, since we did not attempt to sort out false positives detected with {\tt TLS} at quarterly jumps and then rerun {\tt TLS}, we can expect that the TPRs derived with {\tt TLS} in such an iterative manner could actually be higher than shown in Figs.~\ref{fig:TPR_Radius} and \ref{fig:TPR_Mag}. As a consequence, the application of {\tt TLS} on a set of light curves that were properly corrected for quarter jumps can be expected to produce slightly higher planet yields than in Tables~\ref{tab:planet_yield_mag} and \ref{tab:planet_yield_cams}.

{\tt PSLS} (v1.3) includes long-term drift correction. It corrects for the drift in the CCD positions of the stars due to relativistic velocity aberration and satellite motion. It does not currently take into account, however, any detrending of the light curves from jitter, CCD positional changes from thermal trends induced by spacecraft rotation every three months, regular thermal perturbations caused by the daily data transfer, momentum wheel desaturation, residual outliers not detected by the outlier detection algorithms, or the stitching of parts of the light curves between mask updates -- the last of which is irrelevant for P1 sample stars since their photometry will be extracted using a fitting of the point spread function. Although {\tt W{\={o}}tan} can be expected to remove most of these trends in a reasonable manner while preserving transit signatures, the actual data products of the PLATO mission will be subjected to a detailed detrending of systematic effects. In terms of detrending of systematic effects (but not necessarily of astrophysical variability), the real PLATO data will therefore have a somewhat better quality for transit searches than the {\tt PSLS} light curves that we used.

On the down side, our simulations assume near-continuous uptime and uninterrupted functionality of the PLATO satellite. This might be overly optimistic as demonstrated by the Kepler mission, which achieved an average of ${\sim}~90\,\%$ time on target. Unplanned downtimes of PLATO might outbalance the benefits of improved systematic detrending so that our values in Tables~\ref{tab:planet_yield_mag} and \ref{tab:planet_yield_cams} would remain almost unaffected.

We restricted our study to stars with solar-like activity, while the actual stars to be observed in the P1 sample will naturally exhibit a range of activity patterns. An analysis of the first six months of continuous observations of moderately bright stars (Kepler magnitudes $10.5 \leq {\it Kp} \leq 11.5$) from the Kepler primary mission showed that solar-type stars are intrinsically more active than previously thought \citep{2011ApJS..197....6G}, a result later confirmed with the final four years of Kepler observations \citep{2015AJ....150..133G}. Our assumptions of solar-like activity might thus be overly optimistic. This might have a negative effect on the planet yields that we estimate since transit signatures of small planets are harder to find around photometrically more active stars. Simulations of PLATO-like light curves with more realistic intrinsic stellar variability is, in principle, possible with {\tt PSLS}.  For now, {\tt PSLS} requires a user-defined parameterization of stellar activity but future versions are planned to implement empirical descriptions of the magnetic activity to suit PLATO solar-like oscillators \citep{Samadi_2019}. Moreover, rotational modulation caused by spots on the surface of the star are not yet implemented in {\tt PSLS}.

We did not simulate a representative population of 15,000 to 20,000 F5-K7 stars with $m_V\,{\leq}\,11$ as will be observed in PLATO's P1 stellar sample. Instead, we assumed a solar-type star with solar radius and mass, investigated four apparent stellar magnitudes $m_V\,\in\,\{8,9,10,11\}$ for reference simulations with {\tt PSLS}, and weighted the abundance of stars in one-magnitude bins around these reference magnitudes (Sect.~\ref{sec:fov}). There are at least three caveats with this assumption. First, the apparent magnitude distribution of the P1 sample will likely differ from that of field stars, with a drop between $m_V=10$ and $m_V=11$ since the outer, low-camera-count regions of PLATO's field of view are not able to meet the noise limit requirement of 50\,ppm per hour integration (V.~Nascimbeni, priv. comm.). Second, Sun-like stars will likely be underrepresented compared to more early-type stars in the P1 sample. The median stellar radius in the P1 sample will likely be closer to $1.3\,R_\odot$ \citep{2021A&A...653A..98M}, roughly corresponding to spectral type F0 on the main sequence. The HZ will be farther away from the star than in our simulations and the orbital period will be larger than 1\,yr. And third, the P1 sample is not supposed to be evenly distributed over PLATO's NPF and SPF due to the noise limit requirement. Instead, the P1 sample stars will be concentrated in the inner part of the NPF and SPF, where they are covered by 18 or 24 telescopes. In summary, we expect that (1) the apparent magnitude distribution of the P1 sample stars will be skewed more toward brighter stars than based on our Gaia counts of fields stars (Sect.~\ref{sec:fov}), (2) transits of Earth-sized planets in the HZs around P1 sample stars will typically be shallower and have longer orbital periods than the transits around nominal Sun-like stars in our simulations, and (3) the P1 sample stars will preferentially be observed with 18 or 24 cameras. Since points (1) and (3) have an opposite effect on the planet yield to point (2) it is impossible, based on the currently available data, to specify the resulting effect on the actual planet yield presented in this paper.

In all our transit simulations, we assumed a transit impact parameter of zero, that is, that the planet crosses the star along the apparent stellar diameter. In reality, however, the average transit impact parameter for a large sample of transiting planets is $\pi/4 \sim 79\,\%$ of that value \citep{2002ApJ...564..495J}. As a result, we overestimate the average transit duration (and therefore the resulting S/N) around a large sample of Sun-like stars systematically. That said, this effect is mitigated by the longer transit durations expected for HZ planets in the P1 sample, as explained above.

\section{Conclusions}

We have developed a procedure to estimate the yield of Earth-like planets in the HZs around Sun-like stars from the PLATO mission. In brief, we simulated PLATO-like light curves, some of which included transits, with the {\tt PSLS} software, performed a detrending from systematic and astrophysical variability with the {\tt W{\={o}}tan} software, and then searched for the injected signals with the {\tt TLS} search algorithm. We combined our measurements of the TPRs with the expected number of stars in PLATO's P1 stellar sample of bright ($m_V\,{\leq}\,11$) stars and with modern estimates for the occurrence rate of Earth-sized planets in the HZs around Sun-like stars to predict PLATO's exoplanet yield. We investigated the (2\,yr~+~2\,yr) and the (3\,yr~+~1\,yr) observation strategies for PLATO's long-duration observation phase fields.

We find that under the same simulation conditions the (2\,yr~+~2\,yr) observing strategy results in significantly enhanced planet yields compared to the (3\,yr~+~1\,yr) strategy. Details of the exact numbers for both strategies depend on the actual number of stars that will be observed in the P1 sample and on the occurrence rate of small planets in the HZs around Sun-like stars.

Under the assumption of a Sun-like star with low stellar activity, we find that PLATO can detect planets with radii ${\geq}\,1.2\,R_\oplus$ with TPR\,${\sim}100$\,\% in the P1 sample ($m_V~\leq~11$) if two transits can be observed synchronously by 24 PLATO cameras. When a third transit is added under otherwise unchanged conditions, TPR\,${\sim}100$\,\% is achieved for planet as small as Earth. True positive rates of a few percent for planets as small as $0.8\,R_\oplus$ in one-year orbits around bright Sun-like stars from the P1 sample suggest that this is the minimum planet size that can be detected, in some rare and photometrically optimal cases, if two transits are observed. If three transits are available, planets as small as $0.7\,R_\oplus$ may be detectable in rare cases.

Assuming the most unfavorable conditions in our setup with only six PLATO cameras and transits in front of the dimmest Sun-like stars in PLATO's P1 sample ($m_V=11$), TPRs near 100\,\% are nevertheless achieved for planets as small as $1.5\,R_\oplus$ for two transits and as small as $1.3\,R_\oplus$ for three transits around solar-type stars. Again, these estimates all assume low, Sun-like photometric variability.

Using the Sun as a proxy, we predict the detection of between 8 and 34 Earth-like planets in the HZs around F5-K7 main-sequence stars with $0.5\,R_\oplus\,\leq\,R_{\rm p}\,\leq\,1.5\,R_\oplus$. These estimates should be considered an upper limit for several reasons. First, given that PLATO's P1 sample stars will typically be larger than the Sun-like benchmark stars in our simulations \citep{2021A&A...653A..98M}, the resulting transits of Earth-like planets in the HZ will be shallower and less frequent than we simulated. Second, astrophysical false positives, which we neglected, and as yet unknown systematic effects of the PLATO mission might increase the FPR and complicate the identification of genuine transits, in particular given the low number of transits expected for Earth-like planets in the HZs around Sun-like stars. Third, and maybe most importantly, all our estimates are based on simulations of photometrically quiet Sun-like stars, whereas in reality most F5-K7 main-sequence stars are photometrically more variable. On the other hand, a more sophisticated correction of systematic effects and astrophysical variability, more elaborate vetting than a mere SDE-S/N cut, 
a bias of the P1 stellar sample toward bright stars covered with 18 or 24 cameras, and serendipitous discoveries in PLATO's P2--P5 stellar samples could lead to additional discoveries that are not considered in our estimates.

Our results suggest that PLATO can achieve its science objective of finding Earth-like planets in the HZs around Sun-like stars. The prediction of the discovery of 8--34 such worlds means a substantial step forward from the previously available estimates that ranged between 6 and 280. Nevertheless, our new estimates worryingly remind us of the initial predictions for the number of Earth-like planets to be discovered with NASA's Kepler mission, which fluctuated around 50 over the years \citep{1996Ap&SS.241..111B,2016RPPh...79c6901B}. These estimates for Kepler relied on the Sun as a proxy for stellar variability, which turned out to be an overly optimistic approach. Hence, our results require follow-up studies of PLATO's expected planet yield with more realistic stellar variability. If shallow transit detection can be achieved in the presence of significant stellar variability, then our results suggest that PLATO's detections will mean a major contribution to this as yet poorly sampled regime of the exoplanet parameter space with Earth-sized planets in the HZs around solar-type stars.

\begin{acknowledgements}
The authors thank Valerio Nascimbeni, Michael Hippke, Heike Rauer, Juan Cabrera, and Carlos del Burgo for helpful comments on the manuscript. They are also thankful to an anonymous referee for a thorough report. RH acknowledges support from the German Aerospace Agency (Deutsches Zentrum f\"ur Luft- und Raumfahrt) under PLATO Data Center grant 50OO1501. RS acknowledges financial support by the Centre National D'{E}tudes Spatiales (CNES) for the development of {\tt PSLS}. This work presents results from the European Space Agency (ESA) space mission PLATO. The PLATO payload, the PLATO Ground Segment and PLATO data processing are joint developments of ESA and the PLATO Mission Consortium (PMC). Funding for the PMC is provided at national levels, in particular by countries participating in the PLATO Multilateral Agreement (Austria, Belgium, Czech Republic, Denmark, France, Germany, Italy, Netherlands, Portugal, Spain, Sweden, Switzerland, Norway, and United Kingdom) and institutions from Brazil. Members of the PLATO Consortium can be found at \href{https://platomission.com}{https://platomission.com}. The ESA PLATO mission website is \href{https://www.cosmos.esa.int/plato}{https://www.cosmos.esa.int/plato}. We thank the teams working for PLATO for all their work. This work has made use of data from the European Space Agency (ESA) mission {\it Gaia} (\href{https://www.cosmos.esa.int/gaia}{https://www.cosmos.esa.int/gaia}), processed by the {\it Gaia} Data Processing and Analysis Consortium (DPAC, \href{https://www.cosmos.esa.int/web/gaia/dpac/consortium}{https://www.cosmos.esa.int/web/gaia/dpac/consortium}). Funding for the DPAC has been provided by national institutions, in particular the institutions participating in the {\it Gaia} Multilateral Agreement.
\end{acknowledgements}

\bibliographystyle{aa}
\bibliography{literature}

\appendix

\onecolumn

\section{SDE versus S/N for Earth-sized planets and 24 PLATO cameras}
\label{app:SDE-SNR_injections}

\begin{figure*}[h!]
    \centering
    \includegraphics[width=0.497\linewidth]{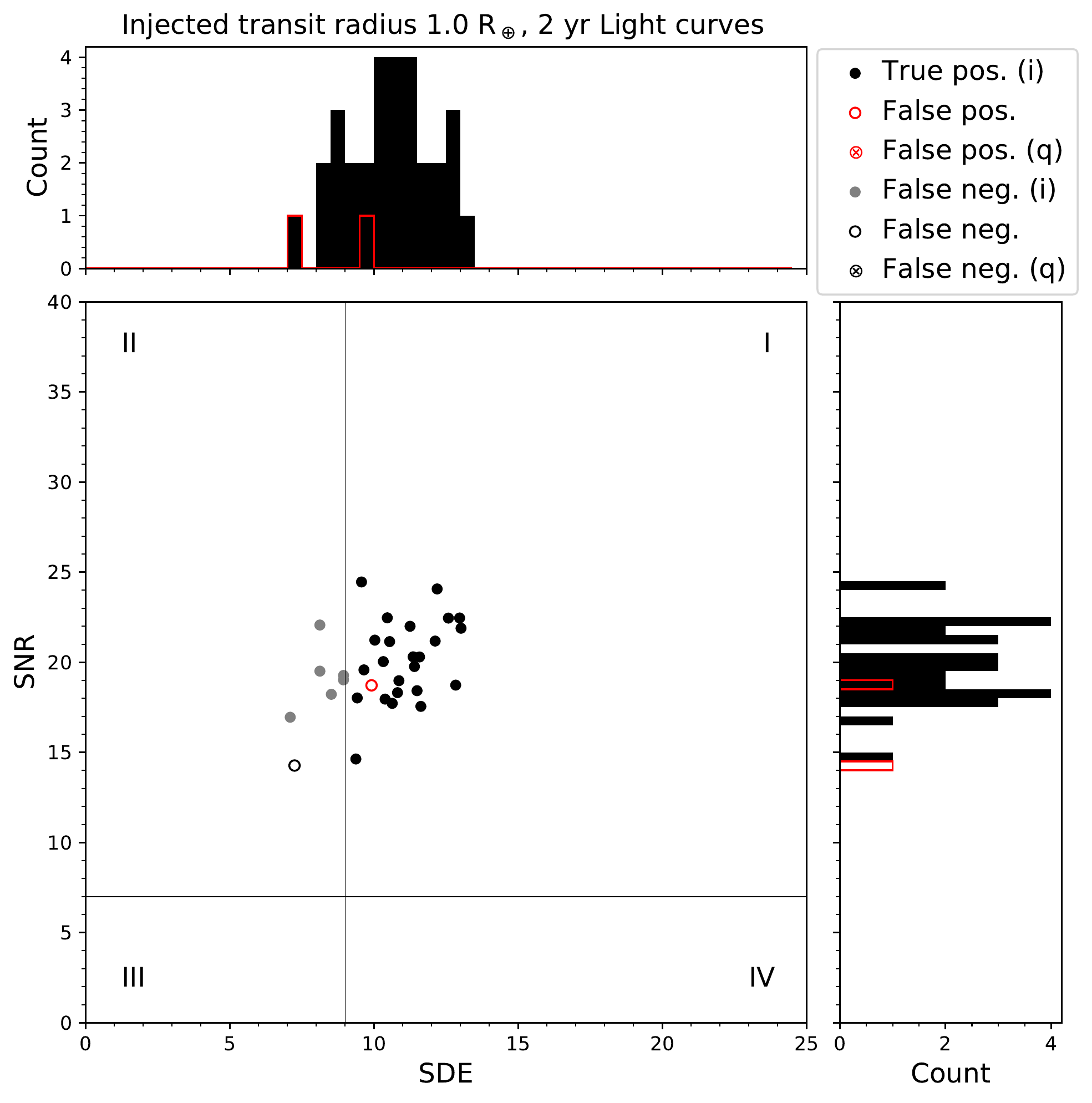}
    \includegraphics[width=0.497\linewidth]{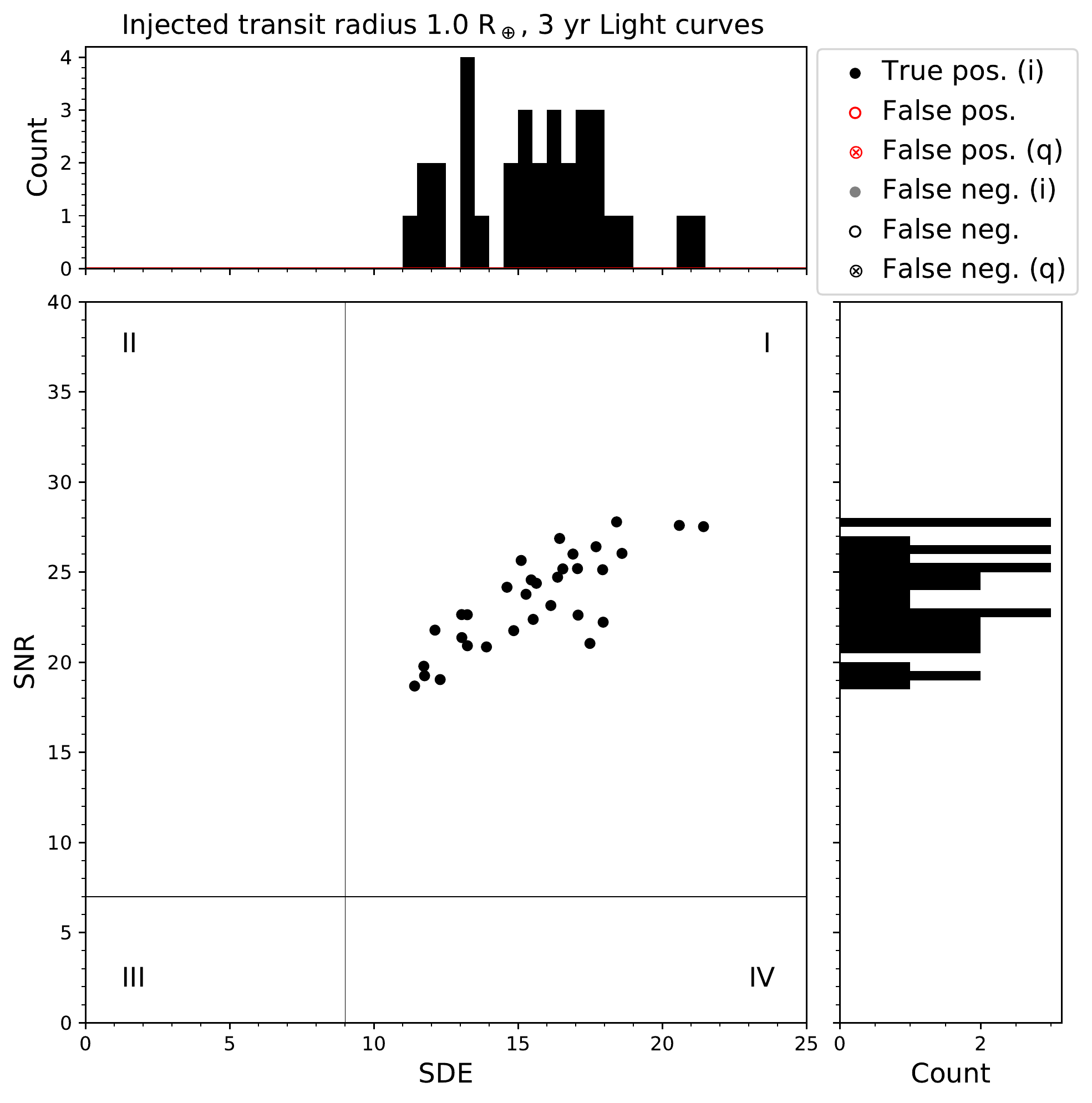}
    \caption{SDE vs. S/N for 100 simulated light curves of an $m_V = 9$ Sun-like star with a $1\,R_\oplus$ transiting planet observed with 24 PLATO cameras. {\it Left}: Each simulated 2\,yr light curve contained two transits. {\it Right}: Each simulated 3\,yr light curve contained three transits.}
    \label{fig:SDE_SNR_10r_2y_3y_mag9}
\end{figure*}

\begin{figure*}[h!]
    \centering
    \includegraphics[width=0.497\linewidth]{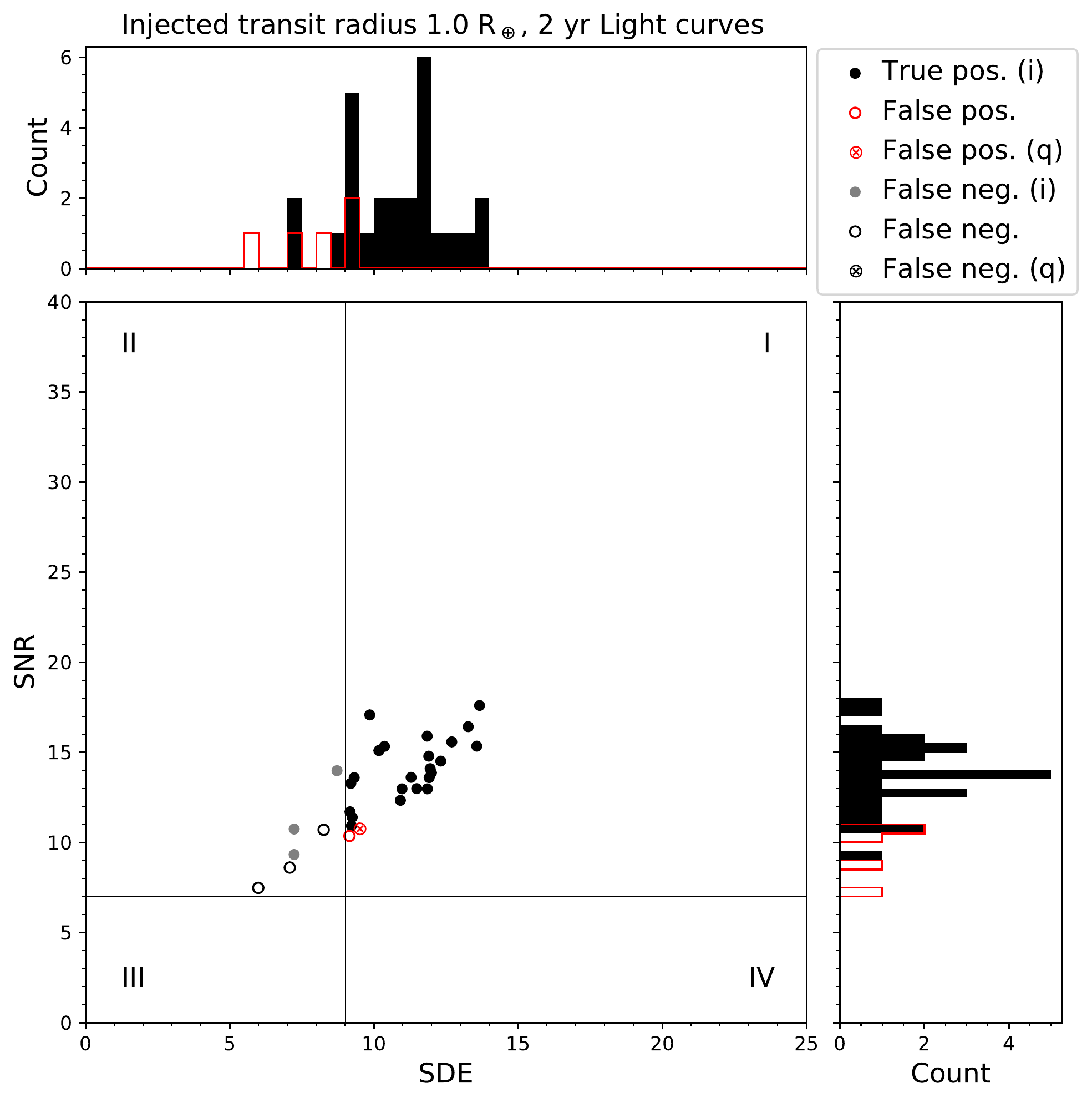}
    \includegraphics[width=0.497\linewidth]{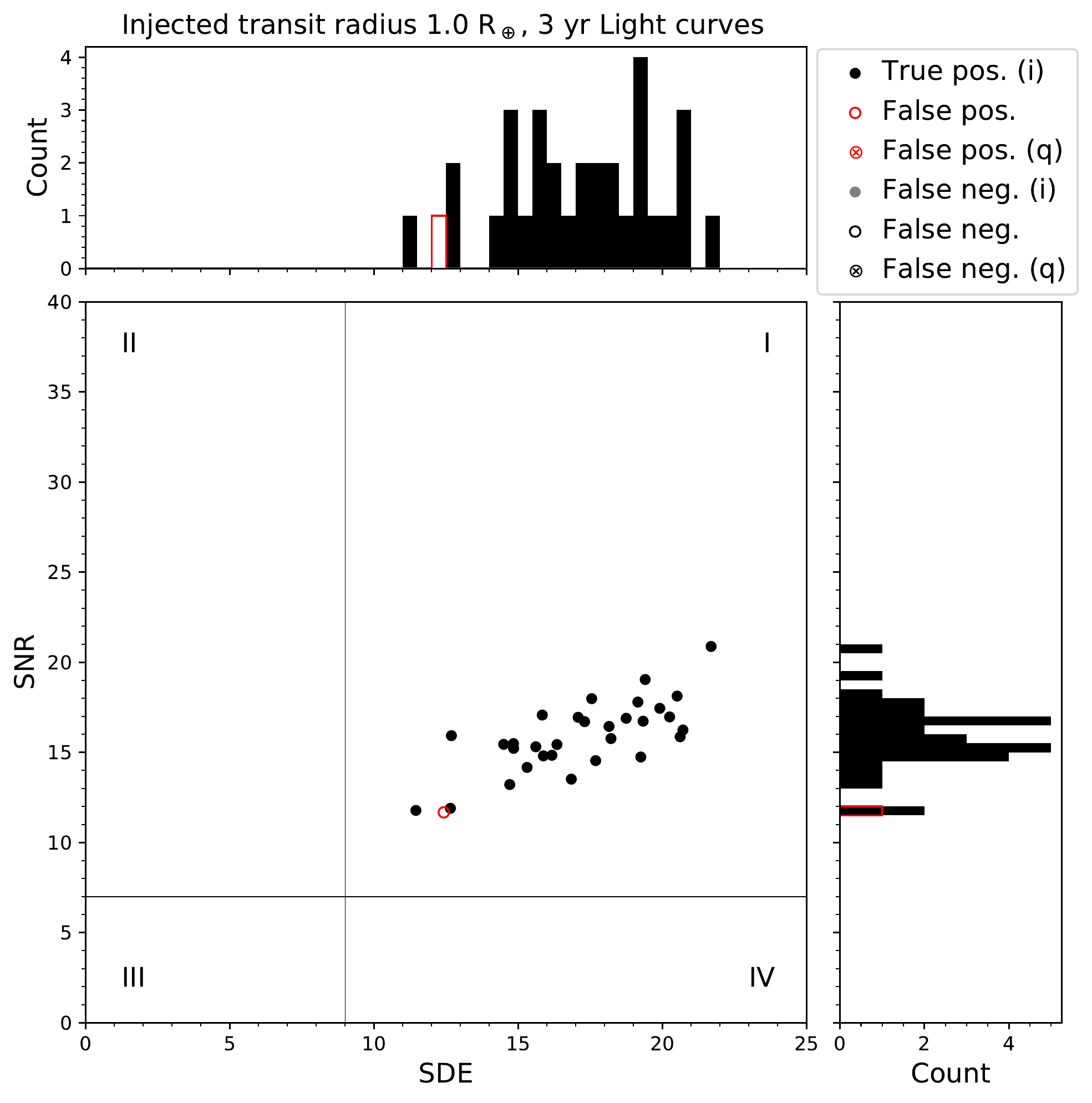}
    \caption{Similar to Fig.~\ref{fig:SDE_SNR_10r_2y_3y_mag9} but for an $m_V = 10$ Sun-like star.}
    \label{fig:SDE_SNR_10r_2y_3y_mag10}
\end{figure*}

\begin{figure*}[h!]
    \centering
    \includegraphics[width=0.497\linewidth]{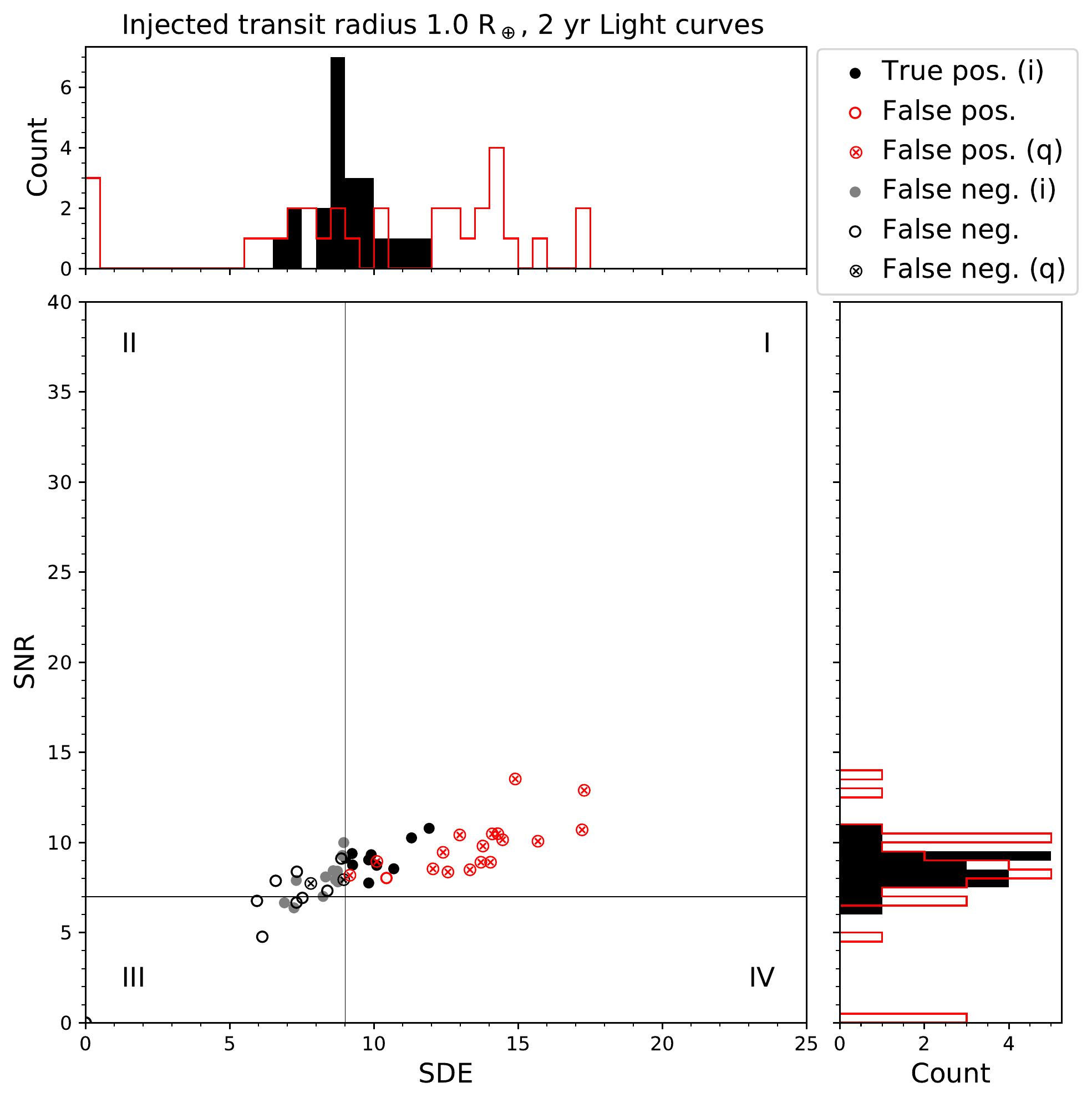}
    \includegraphics[width=0.497\linewidth]{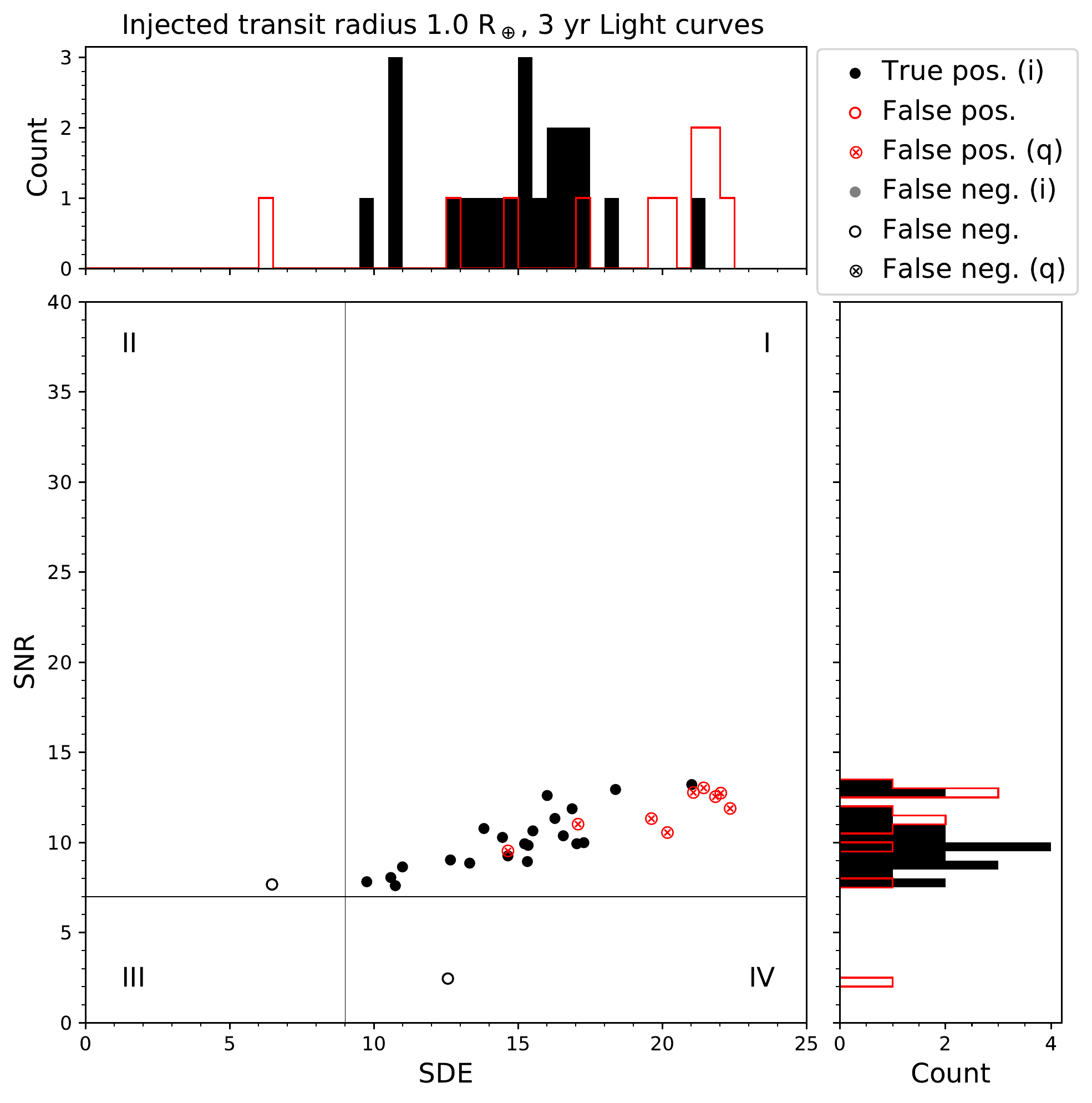}
    \caption{Similar to Fig.~\ref{fig:SDE_SNR_10r_2y_3y_mag9} but for an $m_V = 11$ Sun-like star.}
    \label{fig:SDE_SNR_10r_2y_3y_mag11}
\end{figure*}

As an extension of Fig.~\ref{fig:SDE_SNR_10r_2y_3y} we provide the SDE versus S/N distribution of transiting Earth-like planets around more moderately bright Sun-like stars from the P1 sample. All {\tt PSLS} light curves analyzed for these plots assume observations observed with 24 cameras. Figure~\ref{fig:SDE_SNR_10r_2y_3y_mag9} refers to $m_V = 9$, Fig.~\ref{fig:SDE_SNR_10r_2y_3y_mag10} refers to $m_V = 10$,  and Fig.~\ref{fig:SDE_SNR_10r_2y_3y_mag11} refers to $m_V = 11$ Sun-like stars.


\vspace{7cm}

\section{SDE versus S/N for 24 PLATO cameras without injections}
\label{app:SDE-SNR_noinjections}

\begin{figure*}[h]
    \centering
    \includegraphics[width=0.497\linewidth]{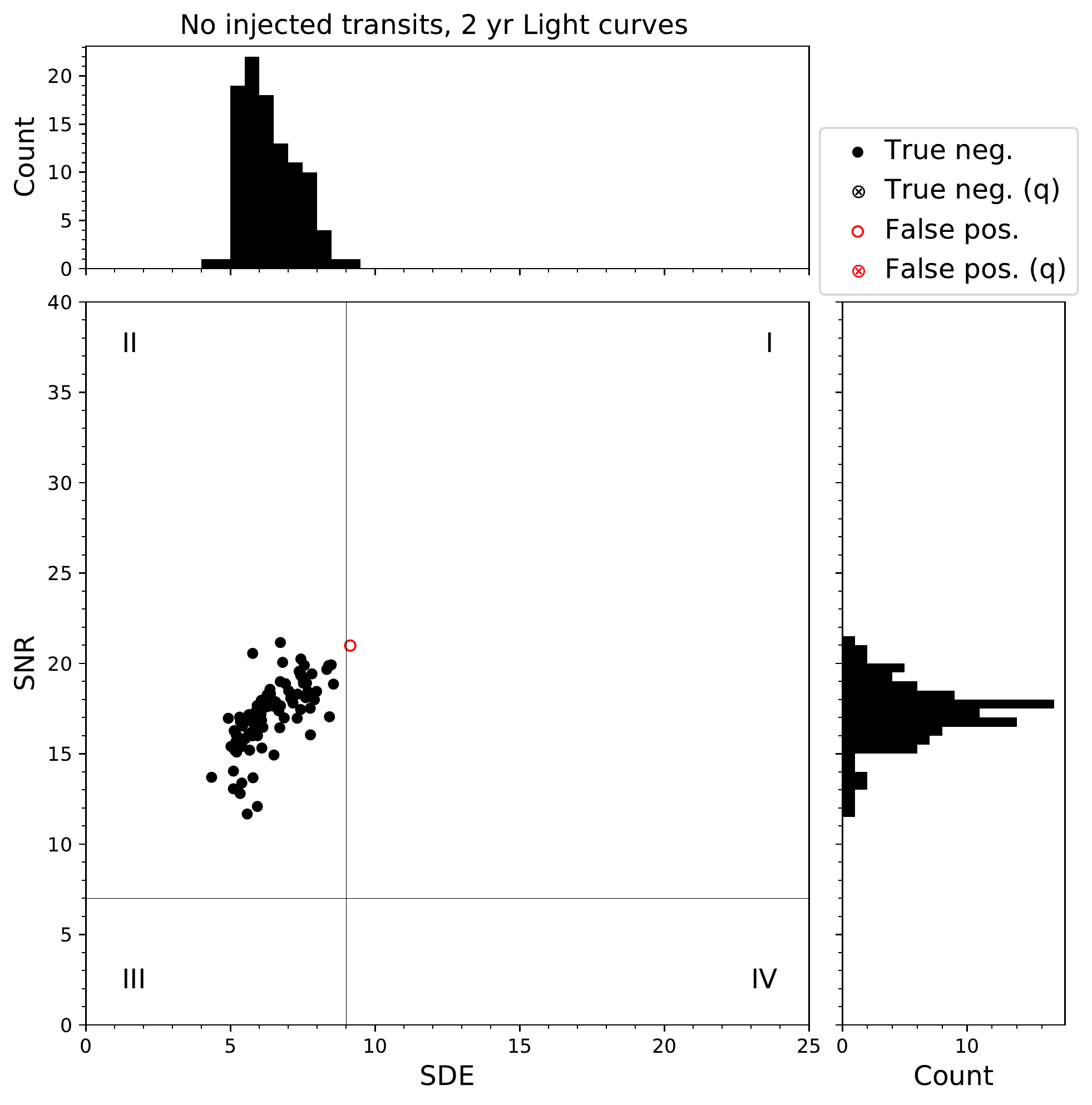}
    \includegraphics[width=0.497\linewidth]{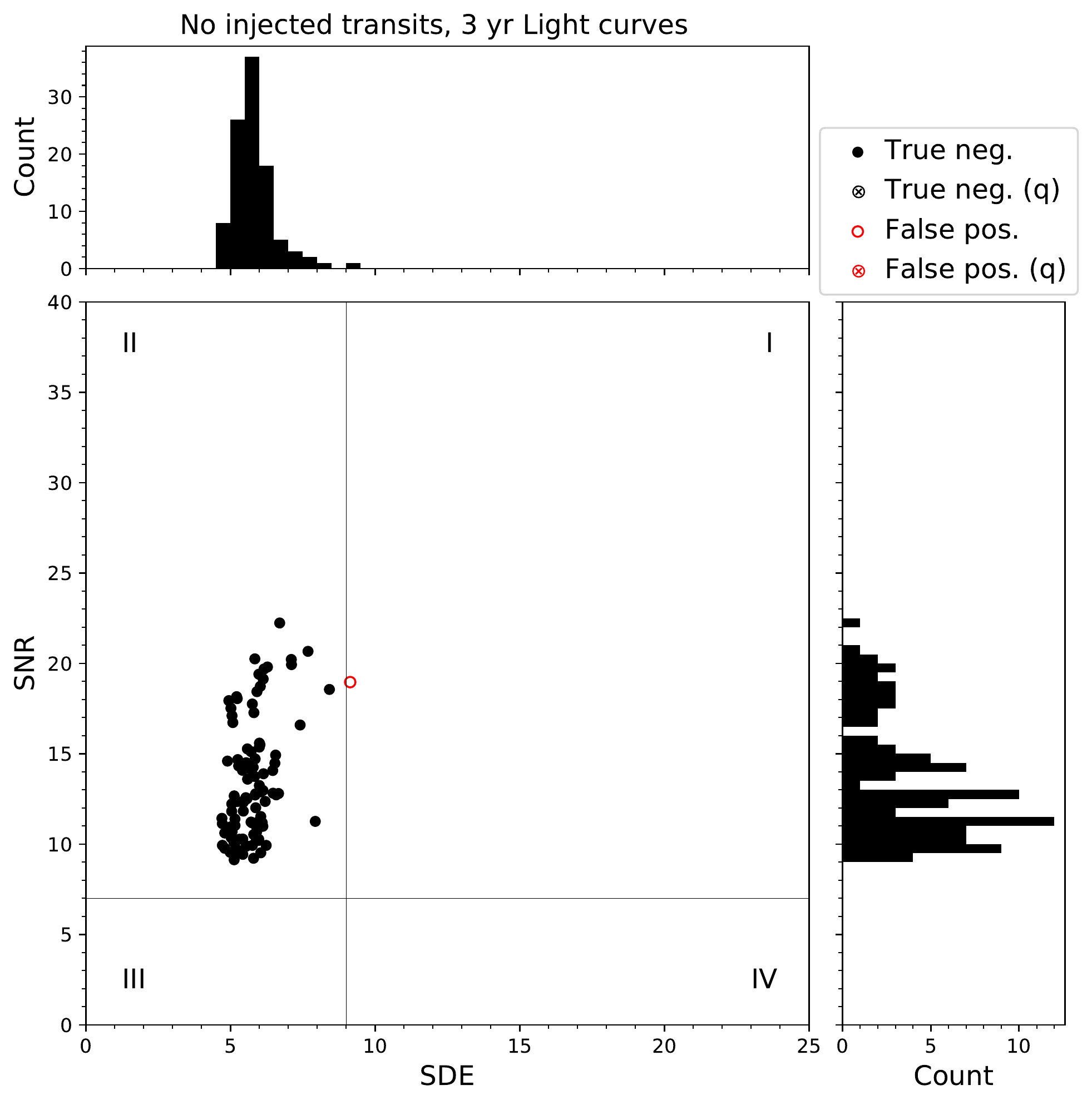}
    \caption{SDE vs. S/N for 100 simulated light curves of an $m_V = 8$ Sun-like star with no injected transiting planet observed with 24 PLATO cameras. {\it Left}: 2\,yr light curves. {\it Right}: 3\,yr light curves.}
    \label{fig:SDE_SNR_no_2y_3y_mag8}
\end{figure*}

\begin{figure*}[h]
    \centering
    \includegraphics[width=0.497\linewidth]{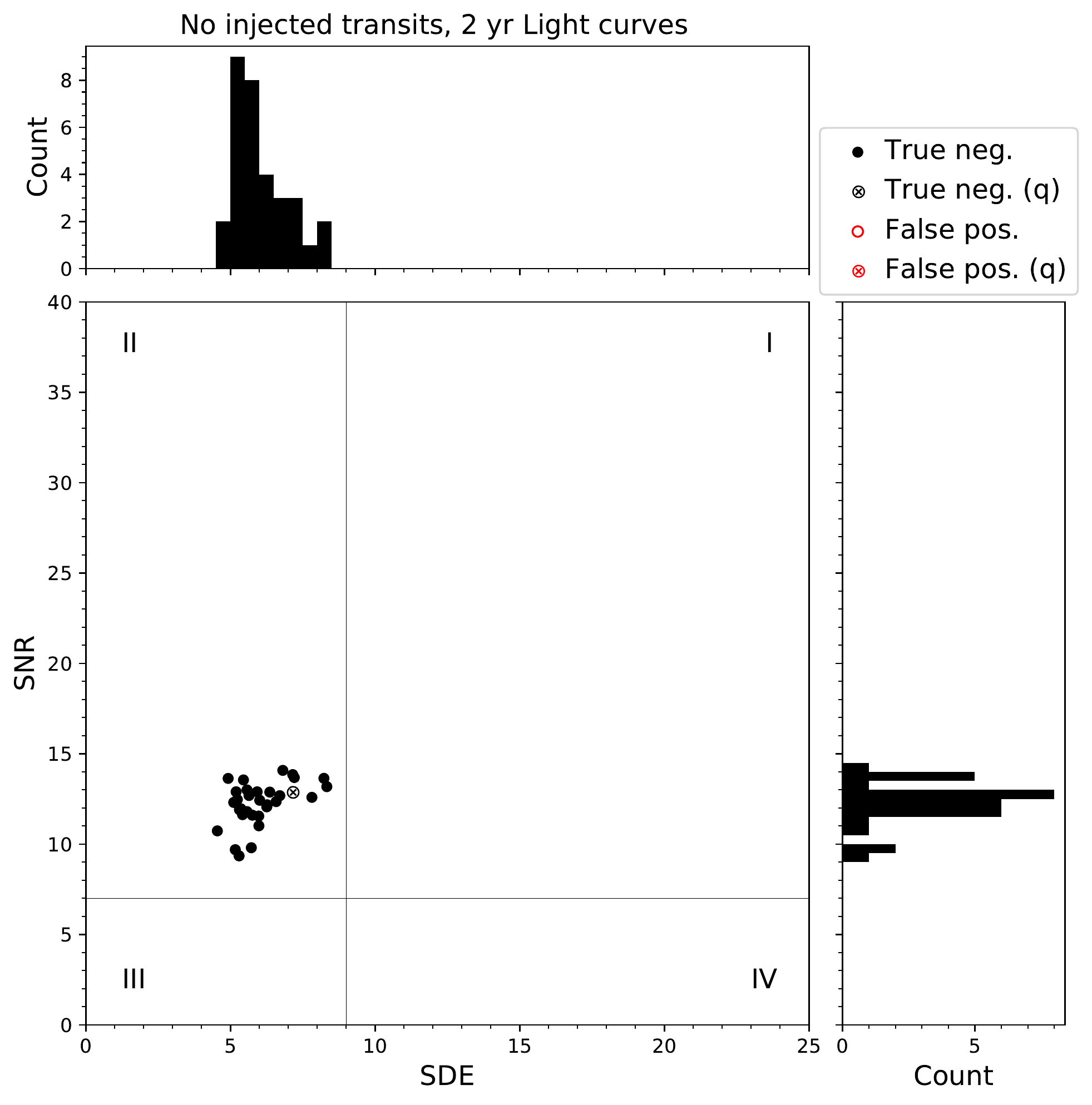}
    \includegraphics[width=0.497\linewidth]{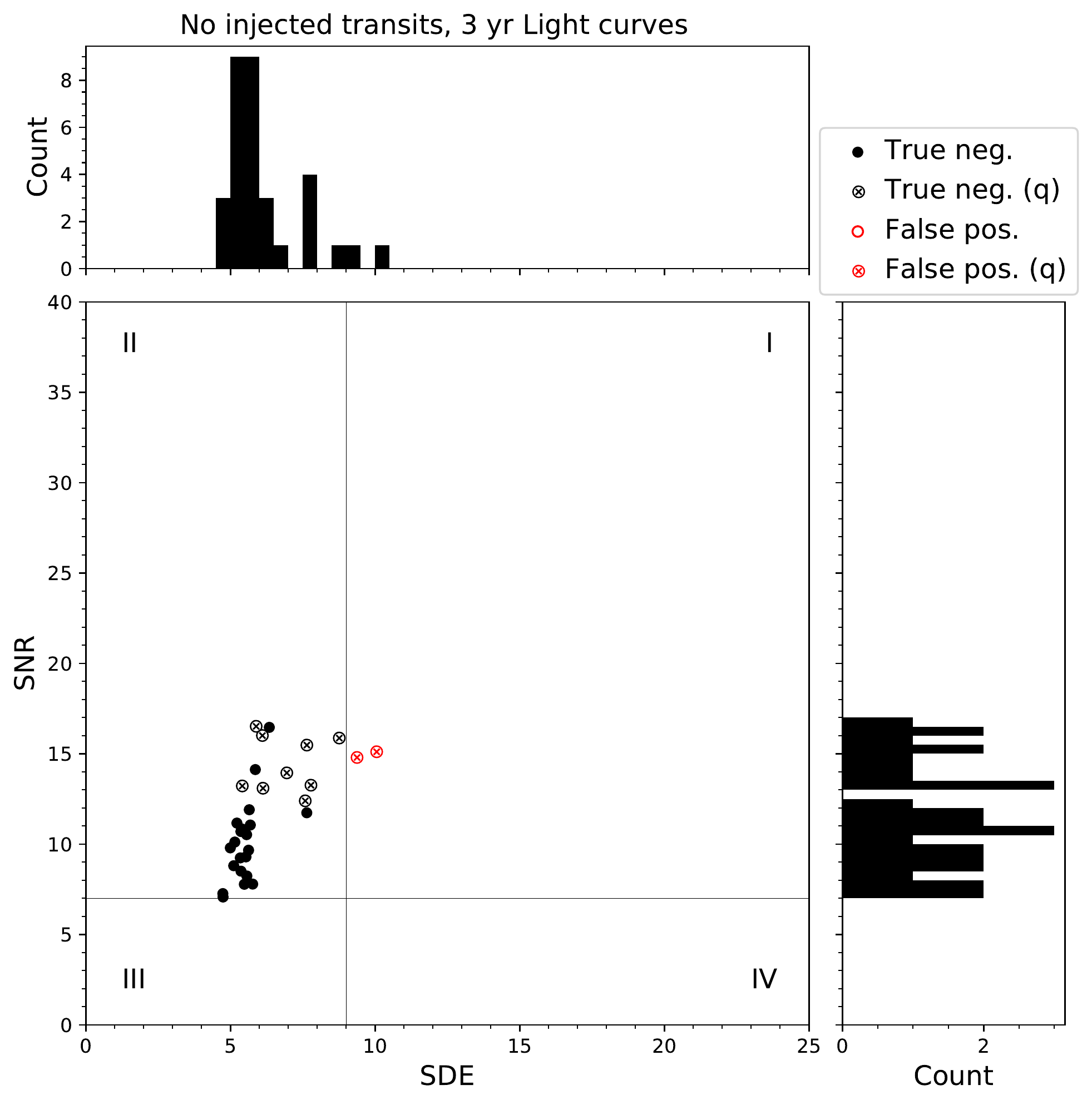}
    \caption{Similar to Fig.~\ref{fig:SDE_SNR_no_2y_3y_mag8} but for an $m_V = 9$ Sun-like star.}
    \label{fig:SDE_SNR_no_2y_3y_mag9}
\end{figure*}

\begin{figure*}[h]
    \centering
    \includegraphics[width=0.497\linewidth]{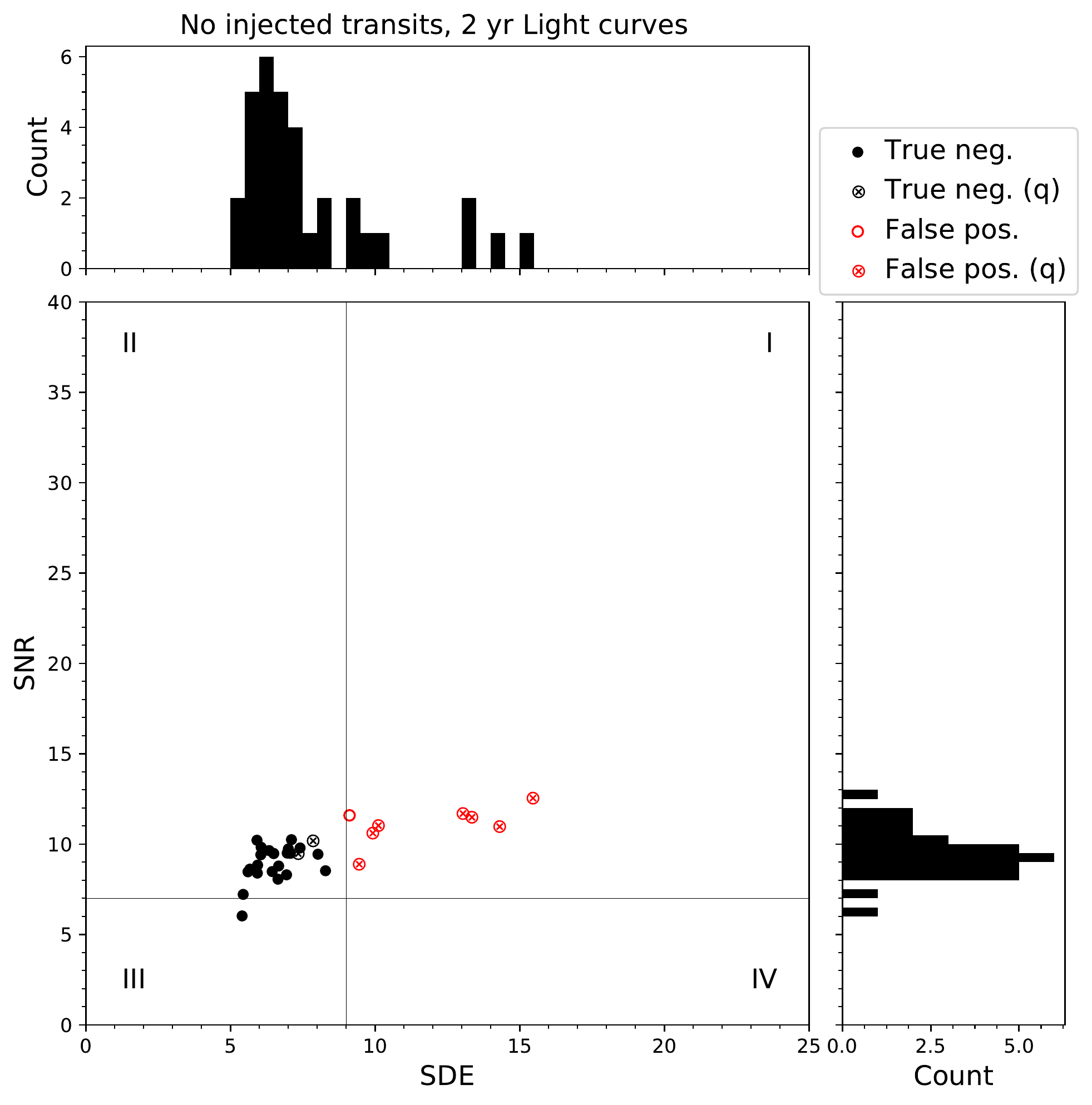}
    \includegraphics[width=0.497\linewidth]{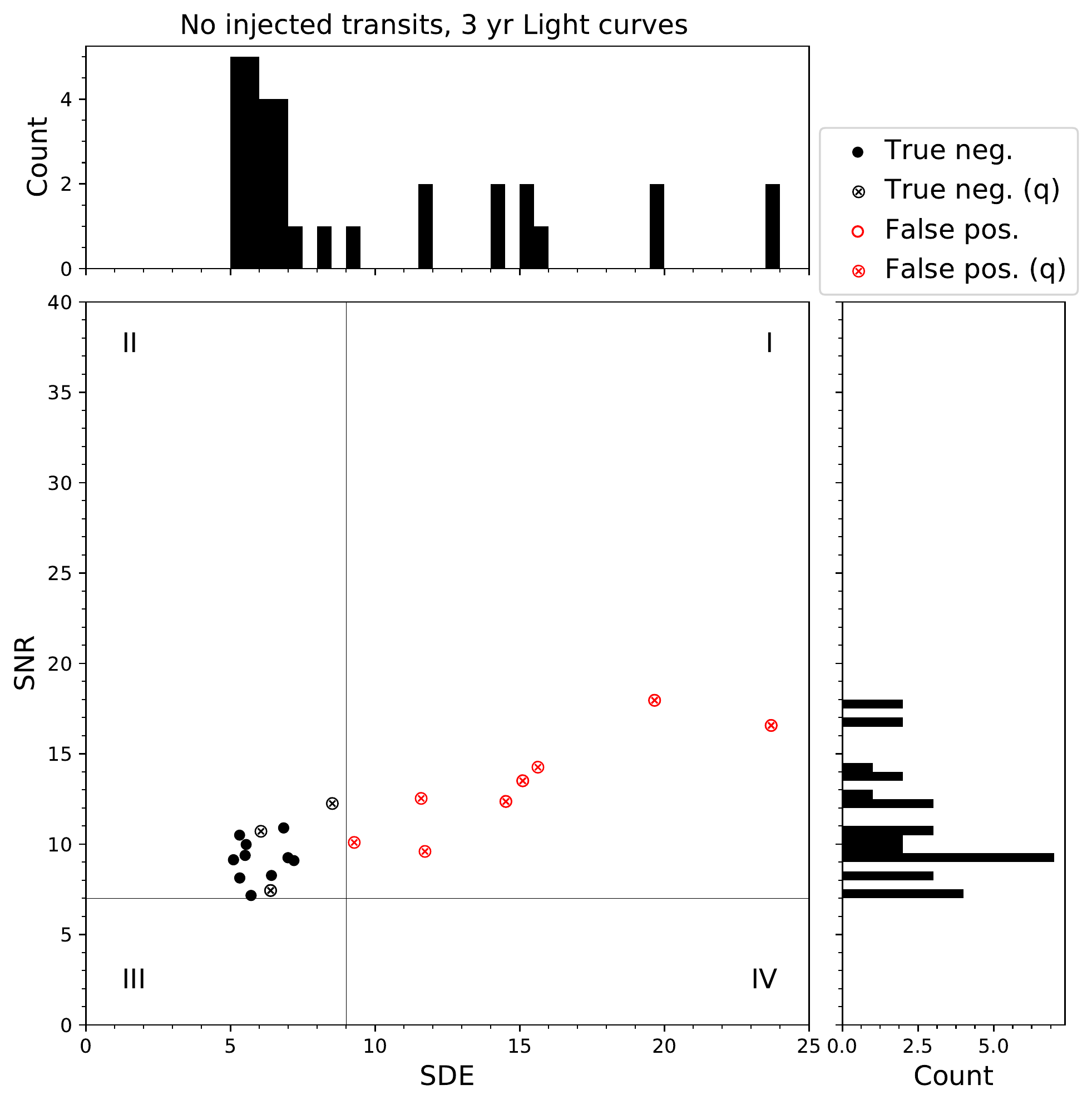}
    \caption{Similar to Fig.~\ref{fig:SDE_SNR_no_2y_3y_mag8} but for an $m_V = 10$ Sun-like star.}
    \label{fig:SDE_SNR_no_2y_3y_mag10}
\end{figure*}

\begin{figure*}[h]
    \centering
    \includegraphics[width=0.497\linewidth]{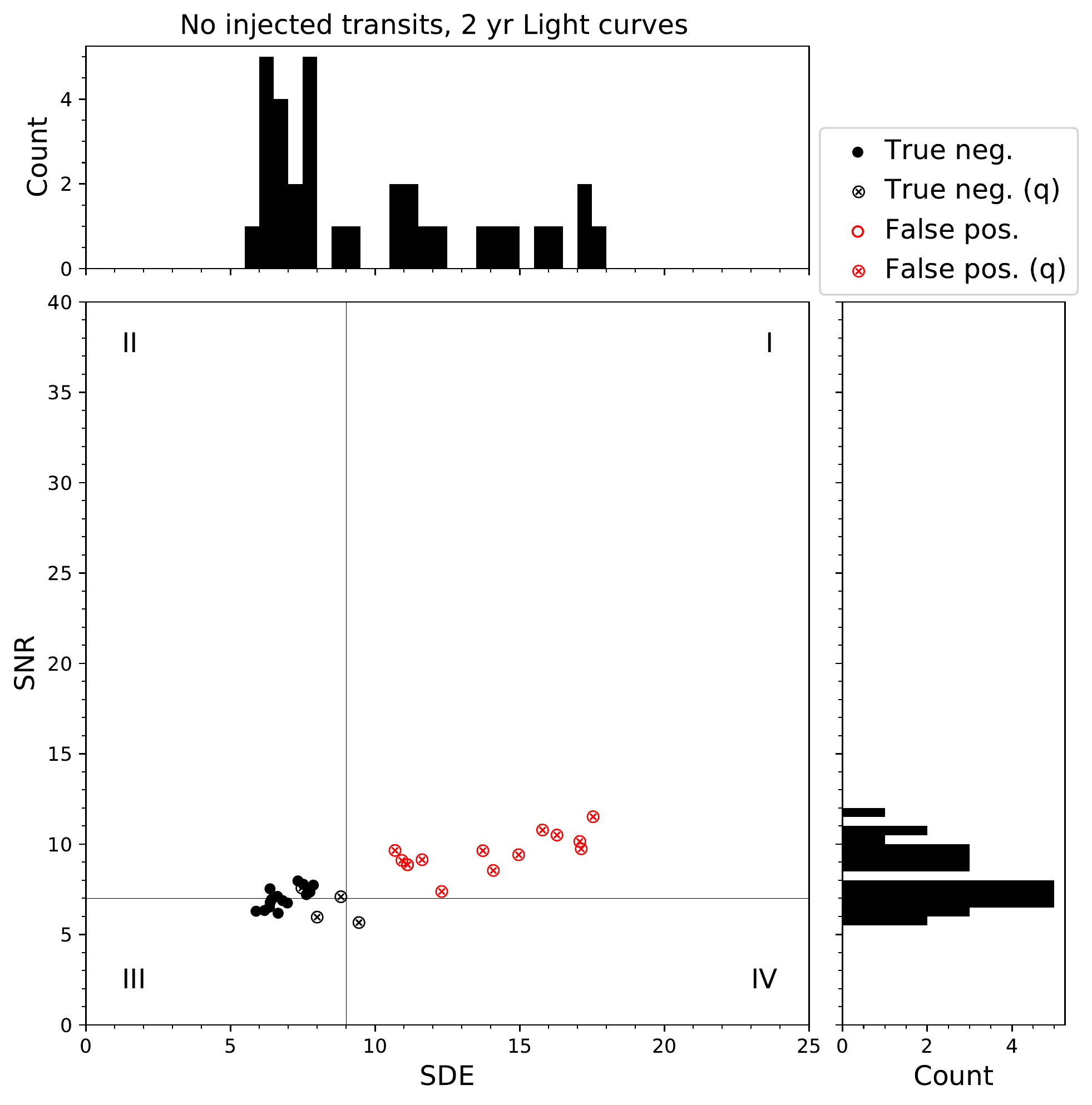}
    \includegraphics[width=0.497\linewidth]{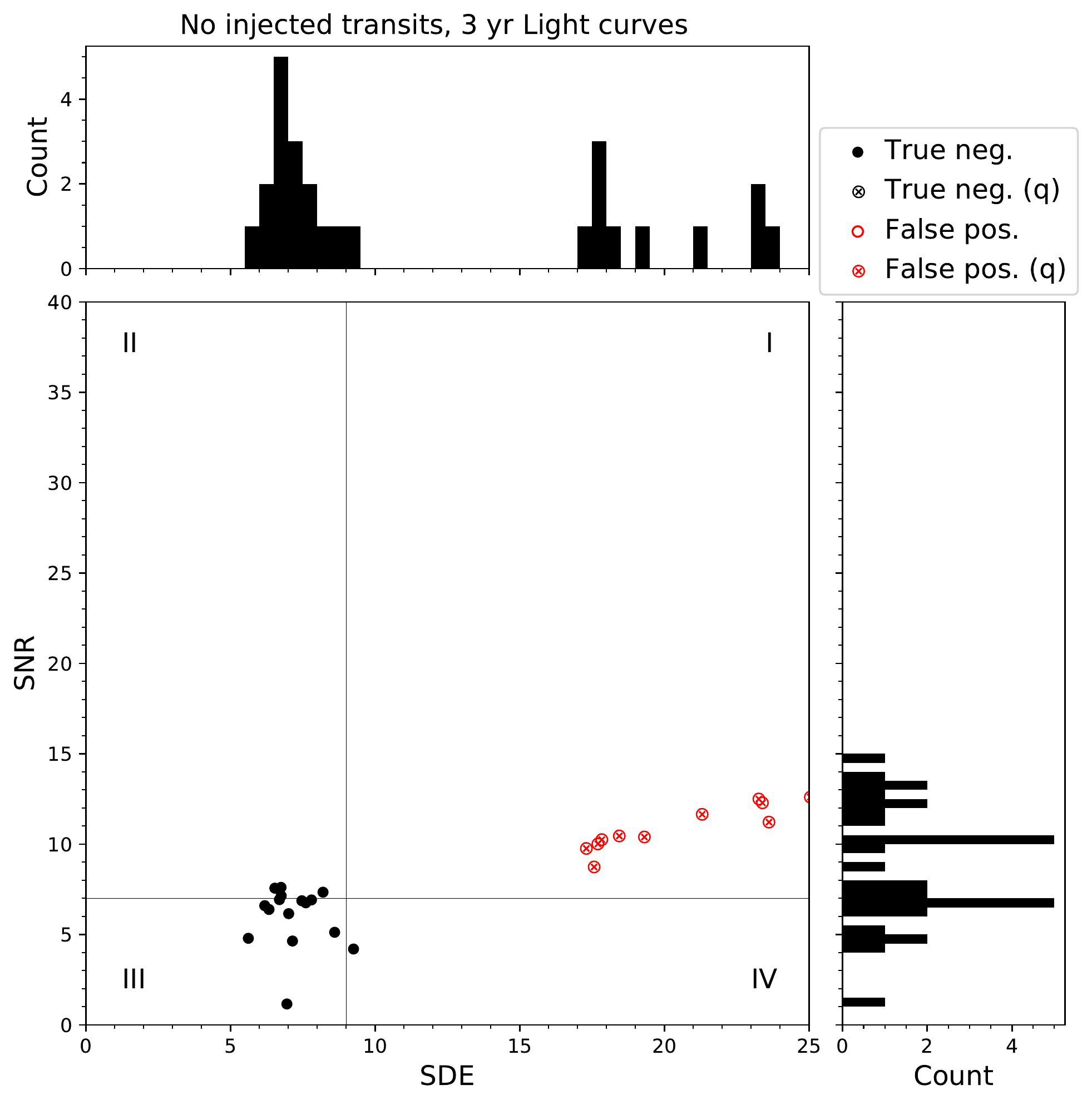}
    \caption{Similar to Fig.~\ref{fig:SDE_SNR_no_2y_3y_mag8} but for an $m_V = 11$ Sun-like star.}
    \label{fig:SDE_SNR_no_2y_3y_mag11}
\end{figure*}

In addition to the SDE versus SDE diagrams for our injection-and-retrieval tests, we also generated similar plots for the control sample of light curves, which did not contain any injected transits. These simulations are key to determining the TNRs and FPRs shown in Fig.~\ref{fig:TNR_Mag}. The following plots are all based on simulated {\tt PSLS} observations with 24 PLATO cameras. Figure~\ref{fig:SDE_SNR_no_2y_3y_mag8} refers to $m_V = 8$, Fig.~\ref{fig:SDE_SNR_no_2y_3y_mag9} refers to $m_V = 9$, Fig.~\ref{fig:SDE_SNR_no_2y_3y_mag10} refers to $m_V = 10$, and Fig.~\ref{fig:SDE_SNR_no_2y_3y_mag11} refers to $m_V = 11$.

\clearpage

\section{Transits in the presence of stellar activity}
\label{app:activity}

Throughout our study we have assumed a Sun-like star with solar activity levels. This assumption had also been made prior to the Kepler mission and unfortunately the analysis of stellar activity levels of the Kepler stars has shown that this assumption was too optimistic. As it turned out, the Sun is a relatively quiet star, from the perspective of photometric variability \citep{2011ApJS..197....6G,2015AJ....150..133G}. The underestimation of stellar variability prior to the Kepler mission is now seen as a main reason for why the mission could not achieve the detection of an Earth-sized planet with three transits, that is, in a one-year orbit around a solar-type star. PLATO will naturally face the same challenges. Although a detailed investigation of the effects of stellar variability on transit detectability is beyond the scope of this manuscript, we have executed a limited qualitative study.

In discussing stellar activity in the context of exoplanet transits, various metrics are in use throughout the community. For example, \citet{2014A&A...572A..34G} measure ${\langle}S_{{\rm ph},k}{\rangle}$, the mean standard deviation in a running window with a width corresponding to $k$ times the rotation period of a given star. As such, ${\langle}S_{{\rm ph},k}{\rangle}$ correlates with the amplitude of the stellar activity component used in the simulations with {\tt PSLS} ($\sigma_{\rm A}$) \citep{Douaglin2018}, which is given in Eq.~(37) in \citet{Samadi_2019}. Beyond stellar activity, ${\langle}S_{{\rm ph},k}{\rangle}$ takes into account any kind of instrumental or systematic error in the light curves and, hence, has the tendency to exceed the stellar activity component. In our analysis we take a similar approach and measure the combined noise level as the standard deviation in a sliding window with a width of 1\,hr ($\sigma_{\rm 1\,hr}$), which is the reference timescale for the computation of the noise-to-signal budgets in the PLATO mission \citep{2014ExA....38..249R}. As an aside, another alternative metric was used by the Kepler mission, which applied the ``combined differential photometric precision'' \citep{2010SPIE.7740E..0DJ}.

To examine transits in light curves with different stellar activity levels, we simulated three stars of solar radius with increasing stellar activity levels and on different characteristic timescales: (1) $\sigma_{\rm A}=40$\,ppm and $\tau_{\rm A}=0.8$\,d; (2) $\sigma_{\rm A}=166$\,ppm and $\tau_{\rm A}=0.5$\,d; and (3) $\sigma_{\rm A}=500$\,ppm and $\tau_{\rm A}=0.3$\,d. The choice of the timescale is motivated by findings of \citet{2011arXiv1104.2185H}, who measured the characteristic timescales for the evolution of stellar activity. As an aside, $\tau_{\rm A}$ used in {\tt PSLS} corresponds to the timescales of \citet{2011arXiv1104.2185H} divided by $2\pi$. As for the activity levels, we refer to \citet{2014A&A...572A..34G}, who determined for the Sun that ${\langle}S_{{\rm ph},k}{\rangle}$, referred to as its photometric magnetic activity level, ranges between 89\,ppm and 259\,ppm with a reference value of 166\,ppm.

\begin{figure*}[h]
    \centering
    \includegraphics[width=0.497\linewidth]{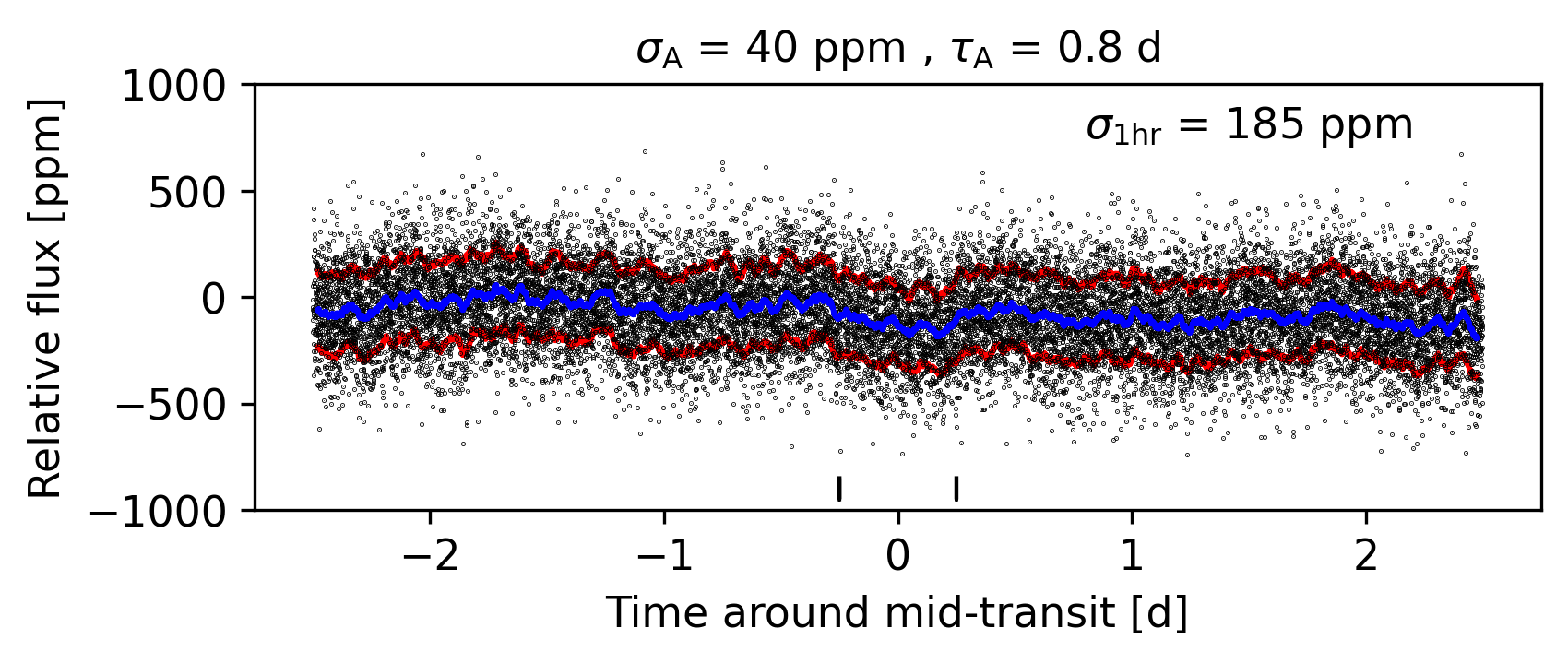}
    \includegraphics[width=0.497\linewidth]{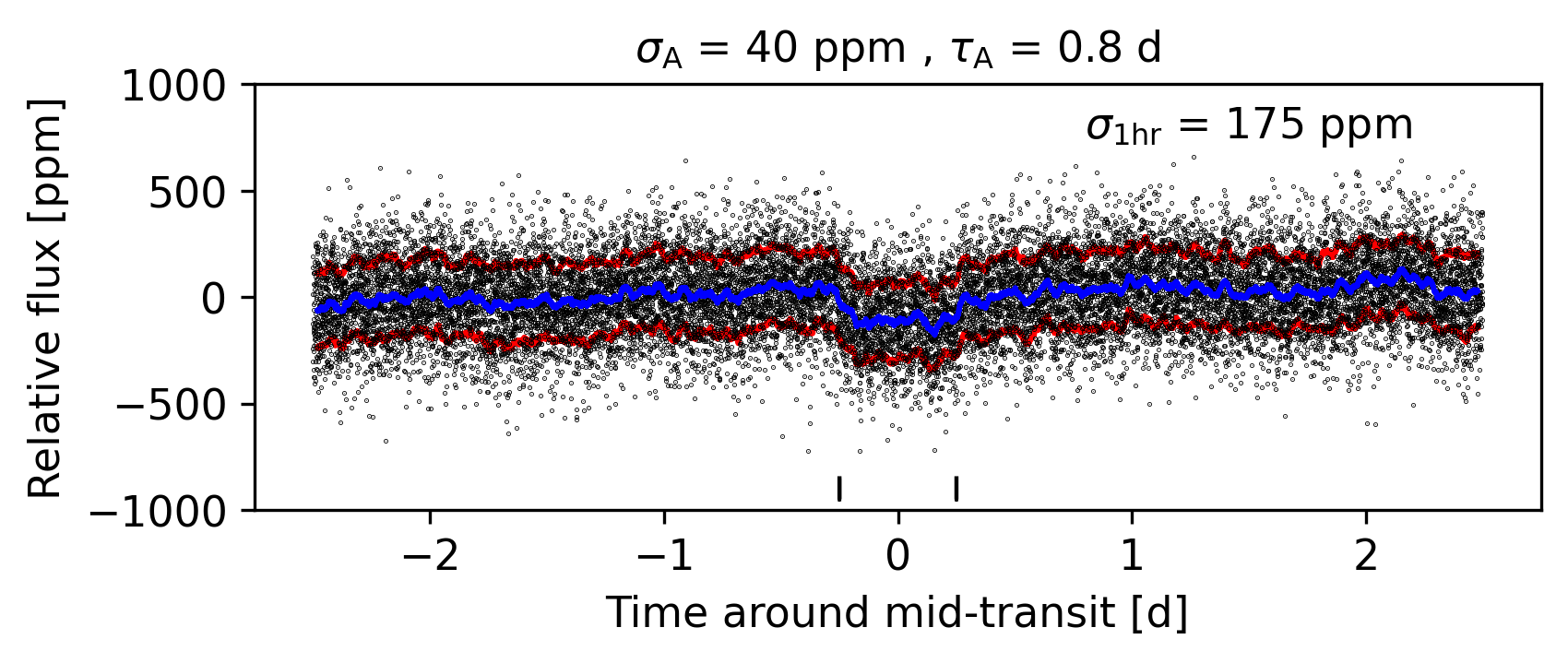}
    \includegraphics[width=0.497\linewidth]{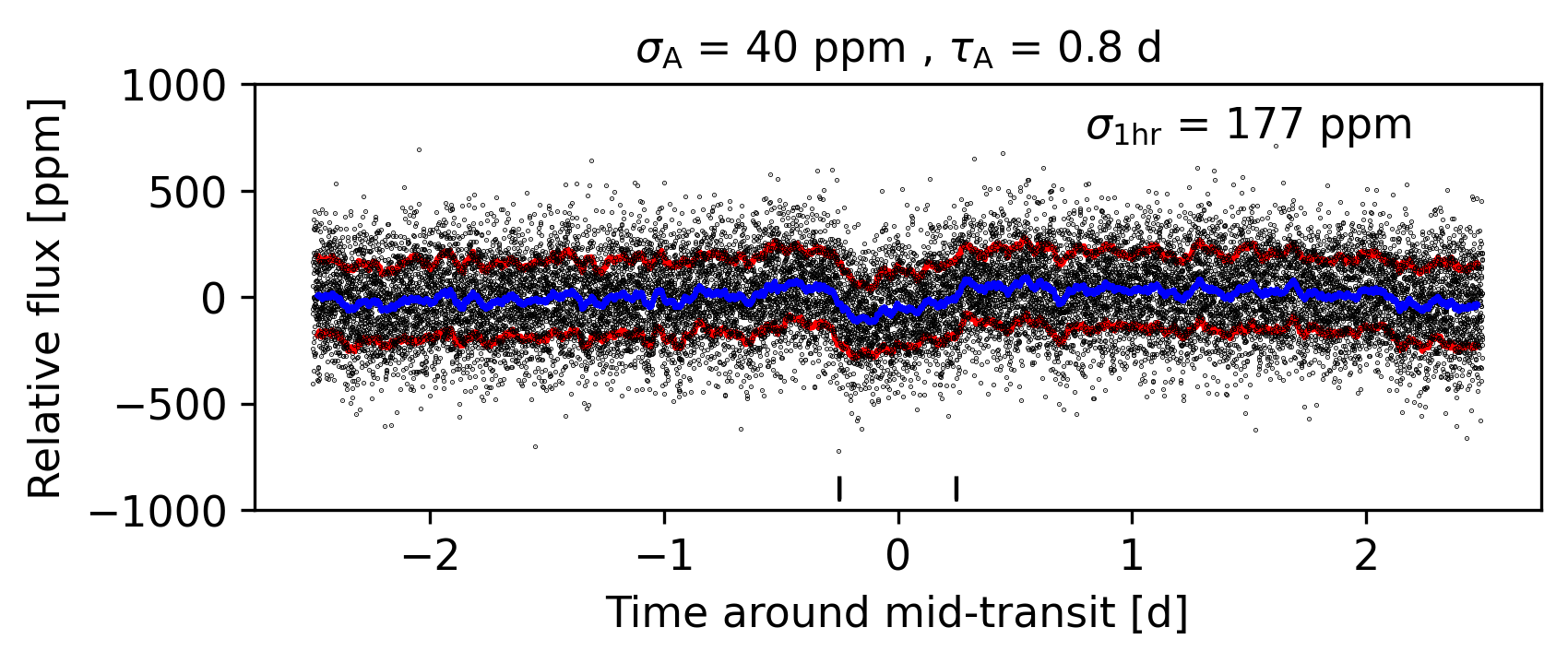}
    \includegraphics[width=0.497\linewidth]{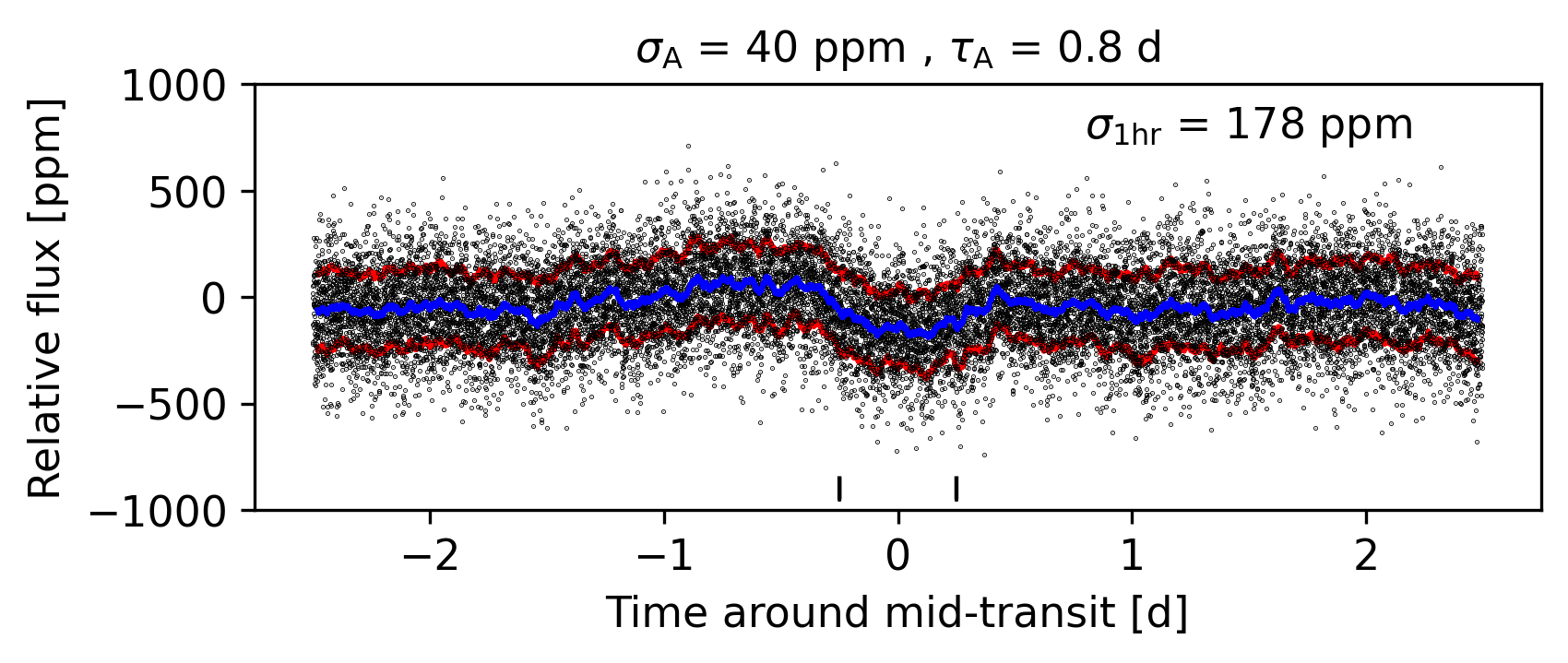}
    \caption{Four example light curves generated with {\tt PSLS} with a transit of an Earth-like planet around a solar-sized star with weak photometric activity ($\sigma_{\rm A}=40$\,ppm, $\tau_{\rm A}=0.8$\,d). The light curves assumed synchronous observations with 24 PLATO cameras and $m_V=9$. The duration of the transit is indicated with two vertical black bars spanning a window of 12\,hr around the mid-transit time.}
    \label{fig:lightcurves_1}
\end{figure*}

\begin{figure*}[h]
    \centering
    \includegraphics[width=0.497\linewidth]{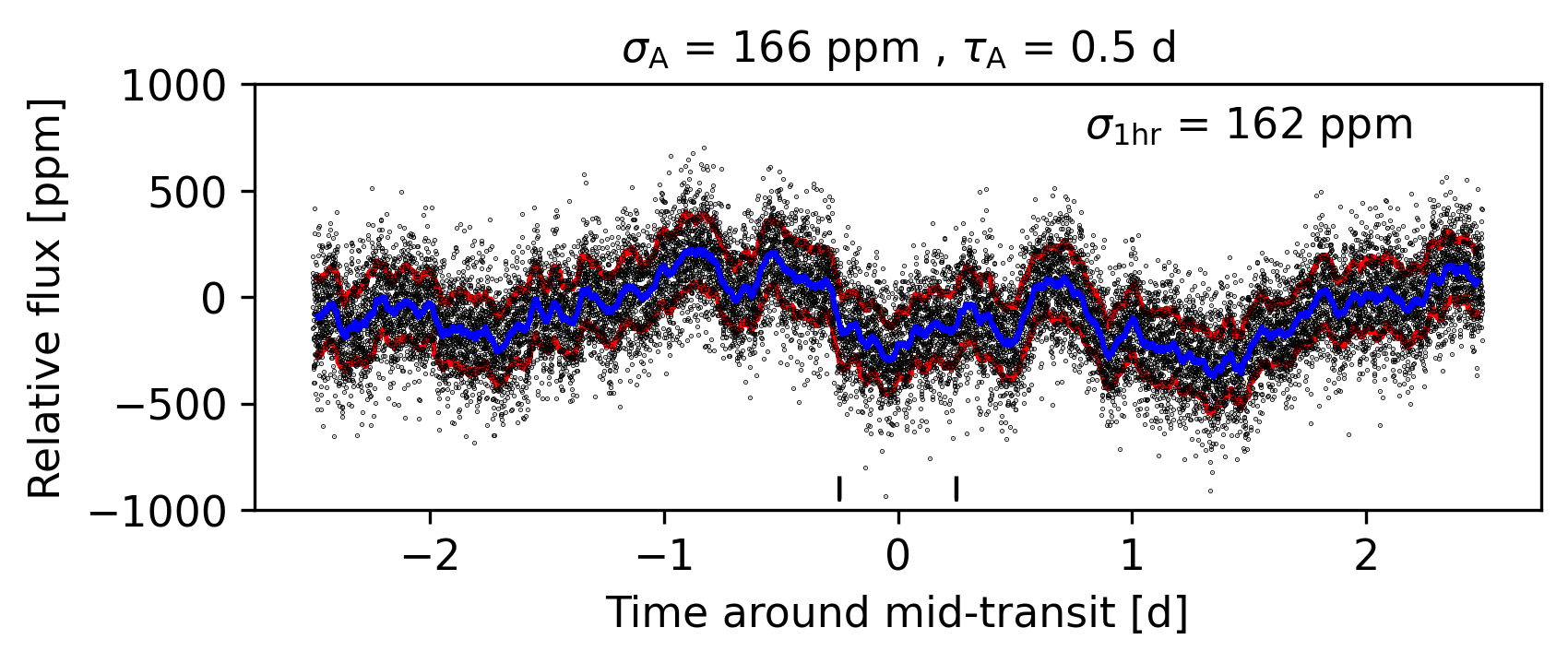}
    \includegraphics[width=0.497\linewidth]{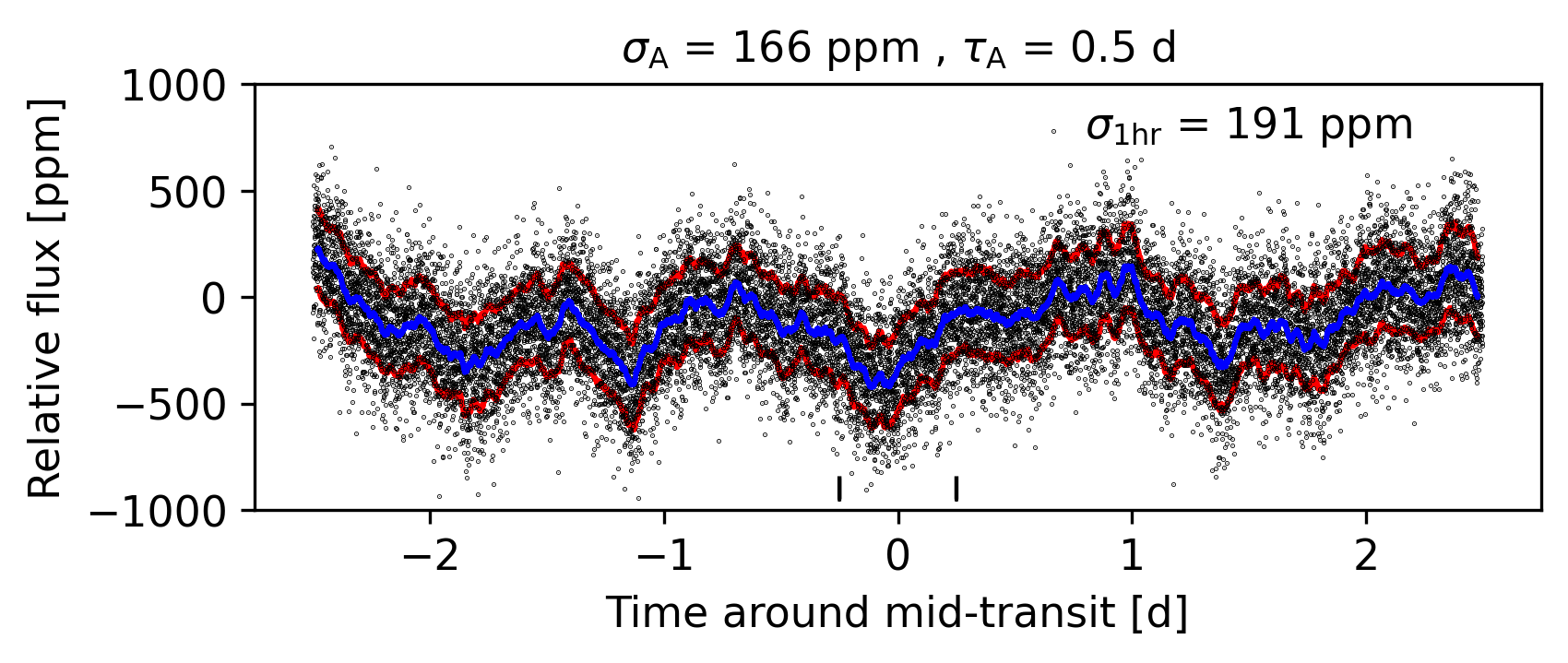}
    \includegraphics[width=0.497\linewidth]{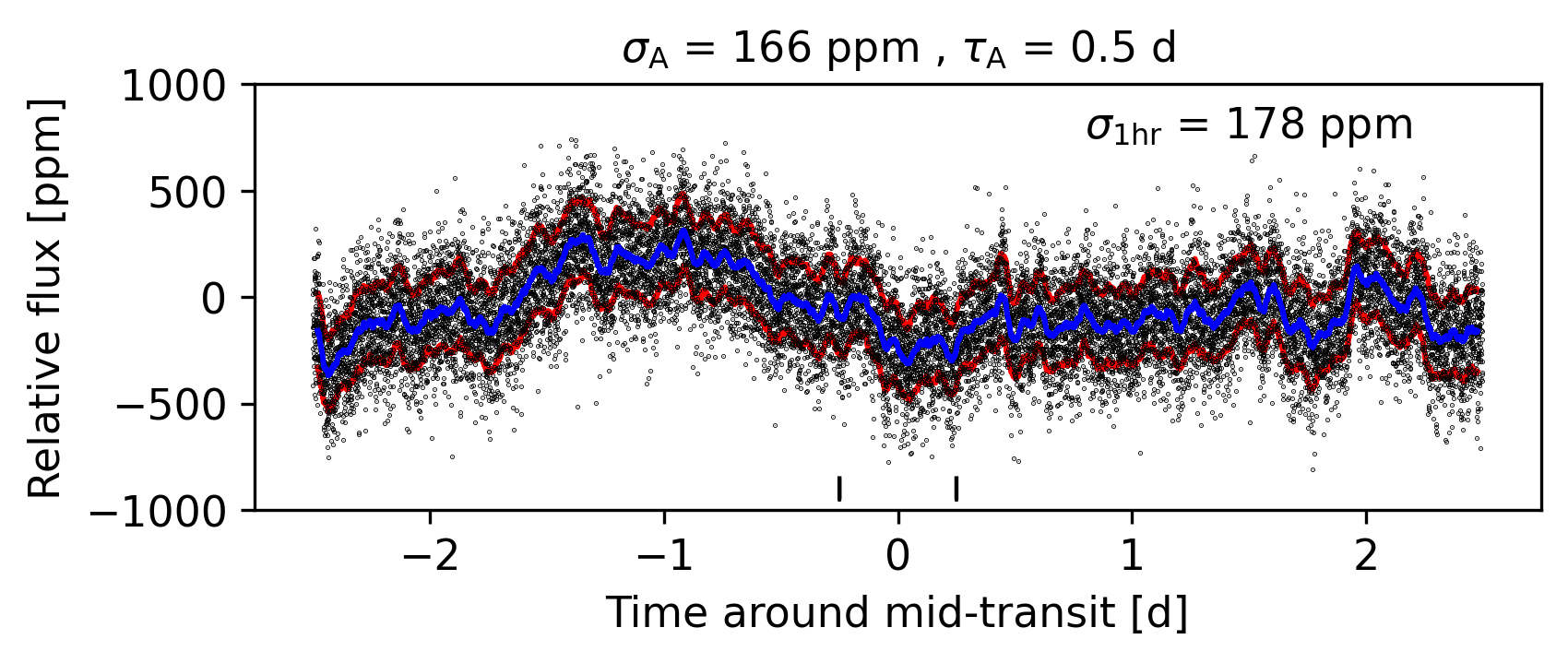}
    \includegraphics[width=0.497\linewidth]{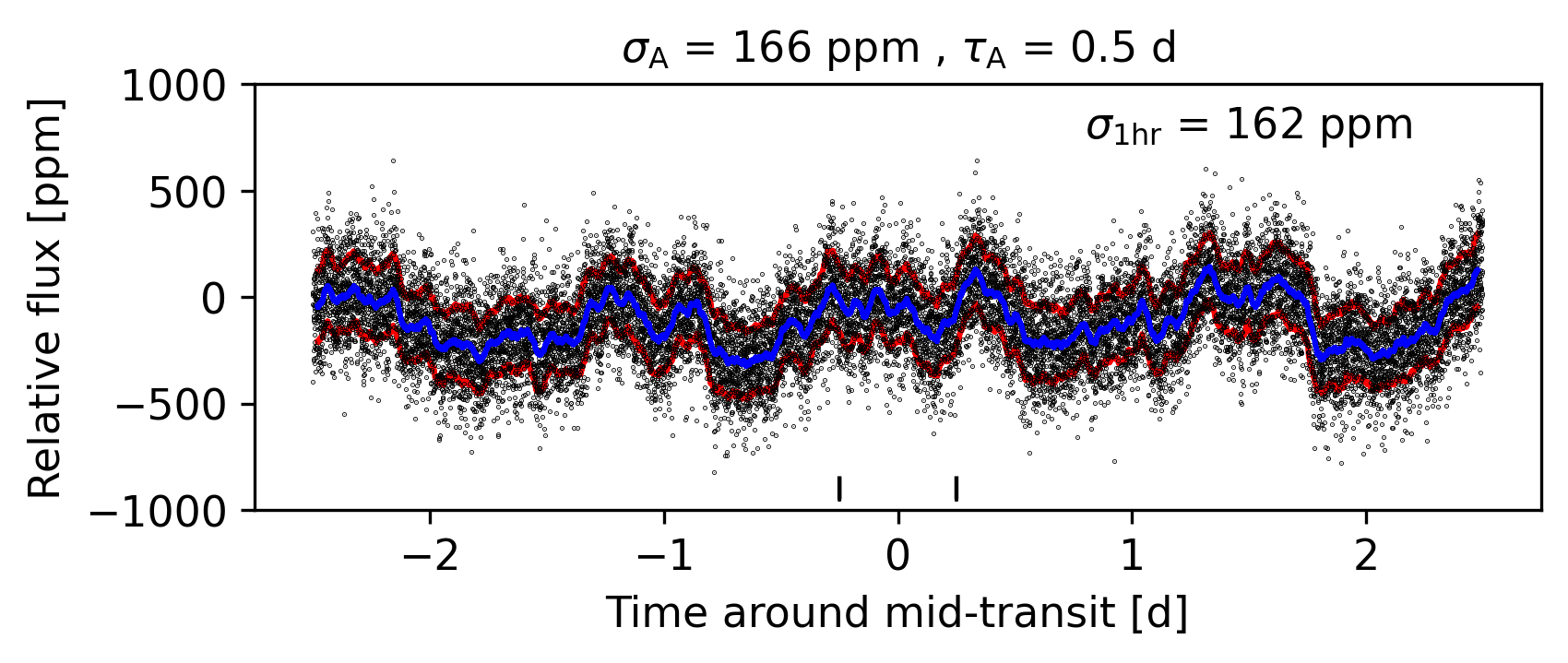}
    \caption{Same as Fig.~\ref{fig:lightcurves_1} but for a moderately active star ($\sigma_{\rm A}=166$\,ppm, $\tau_{\rm A}=0.5$\,d).}
    \label{fig:lightcurves_2}
\end{figure*}

\begin{figure*}[h]
    \centering
    \includegraphics[width=0.497\linewidth]{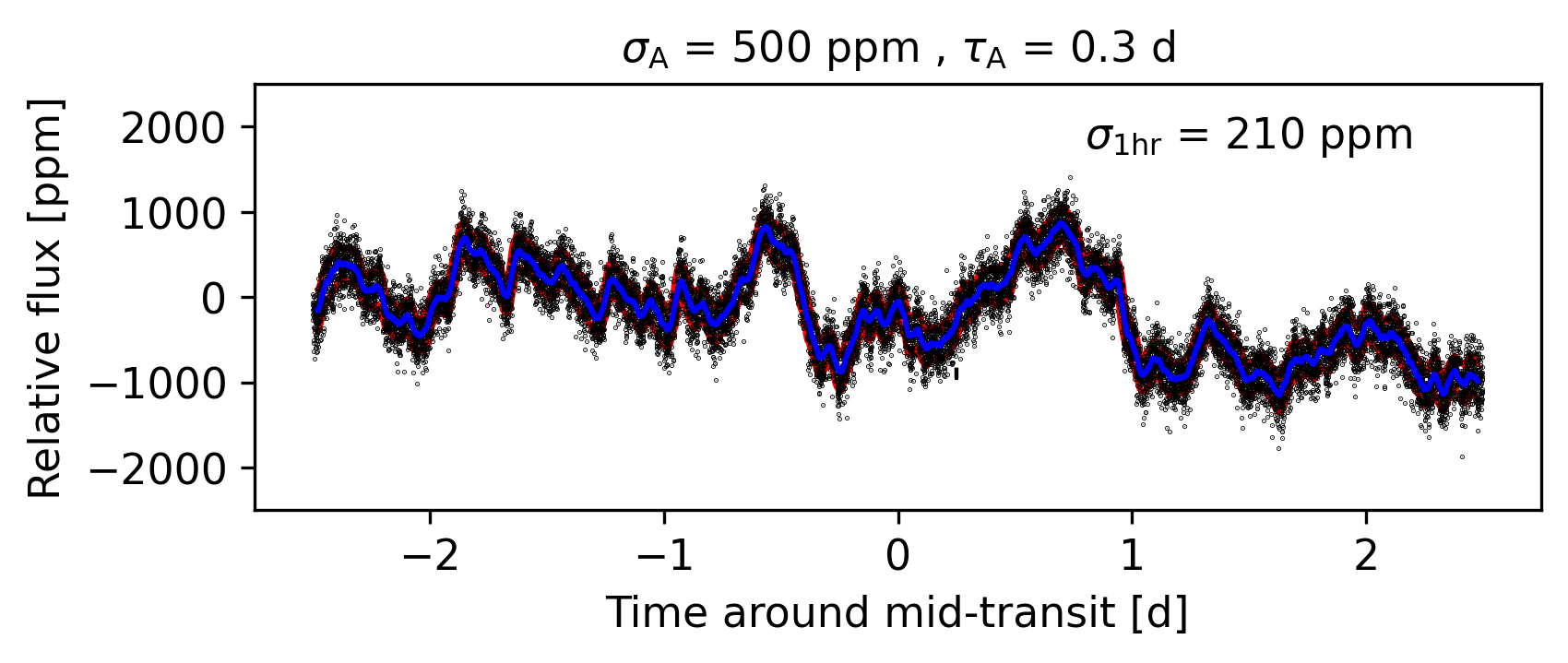}
    \includegraphics[width=0.497\linewidth]{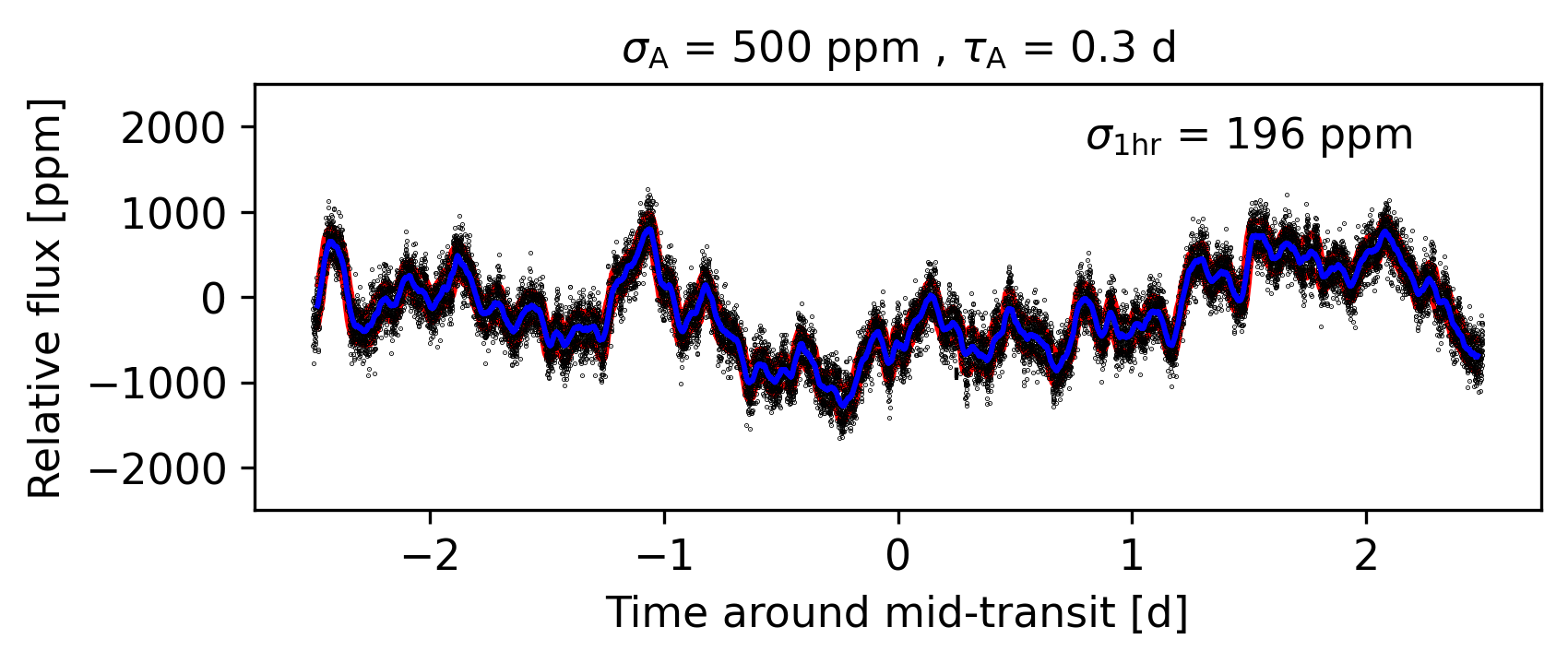}
    \includegraphics[width=0.497\linewidth]{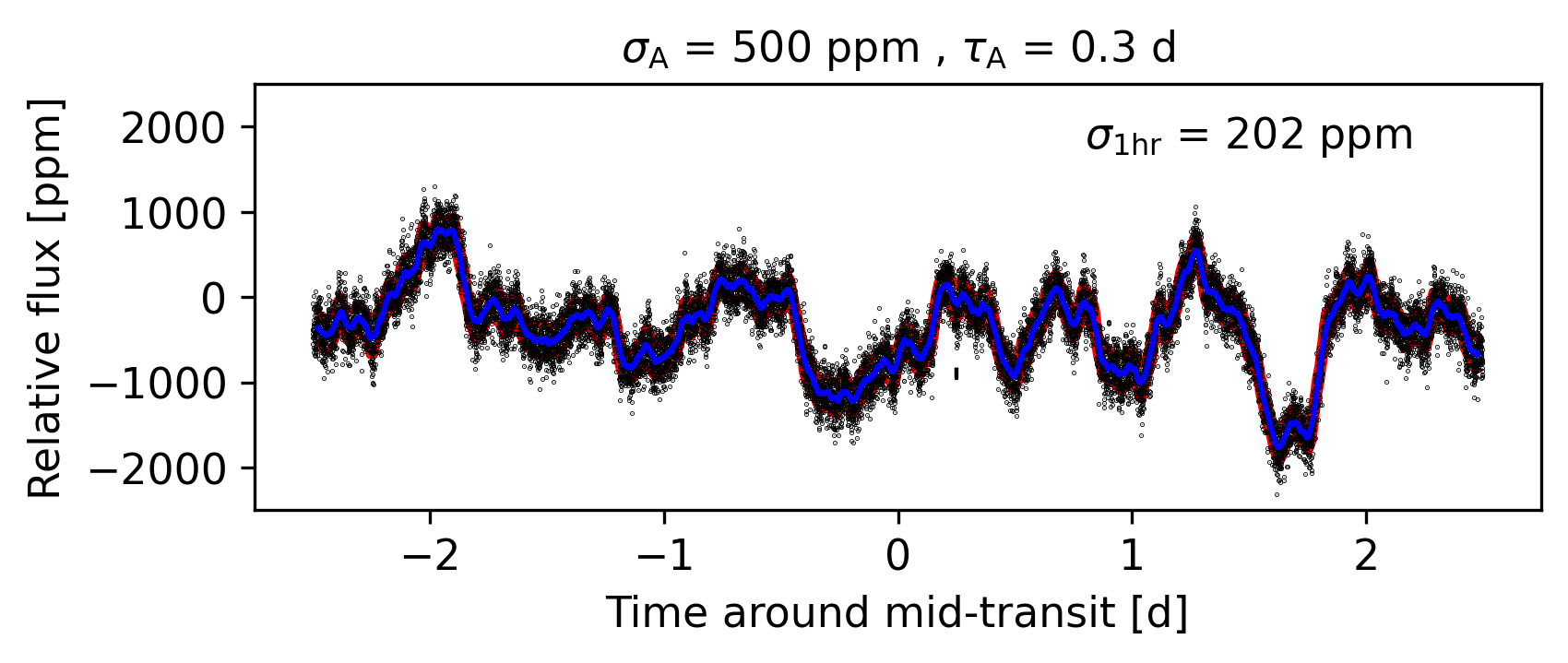}
    \includegraphics[width=0.497\linewidth]{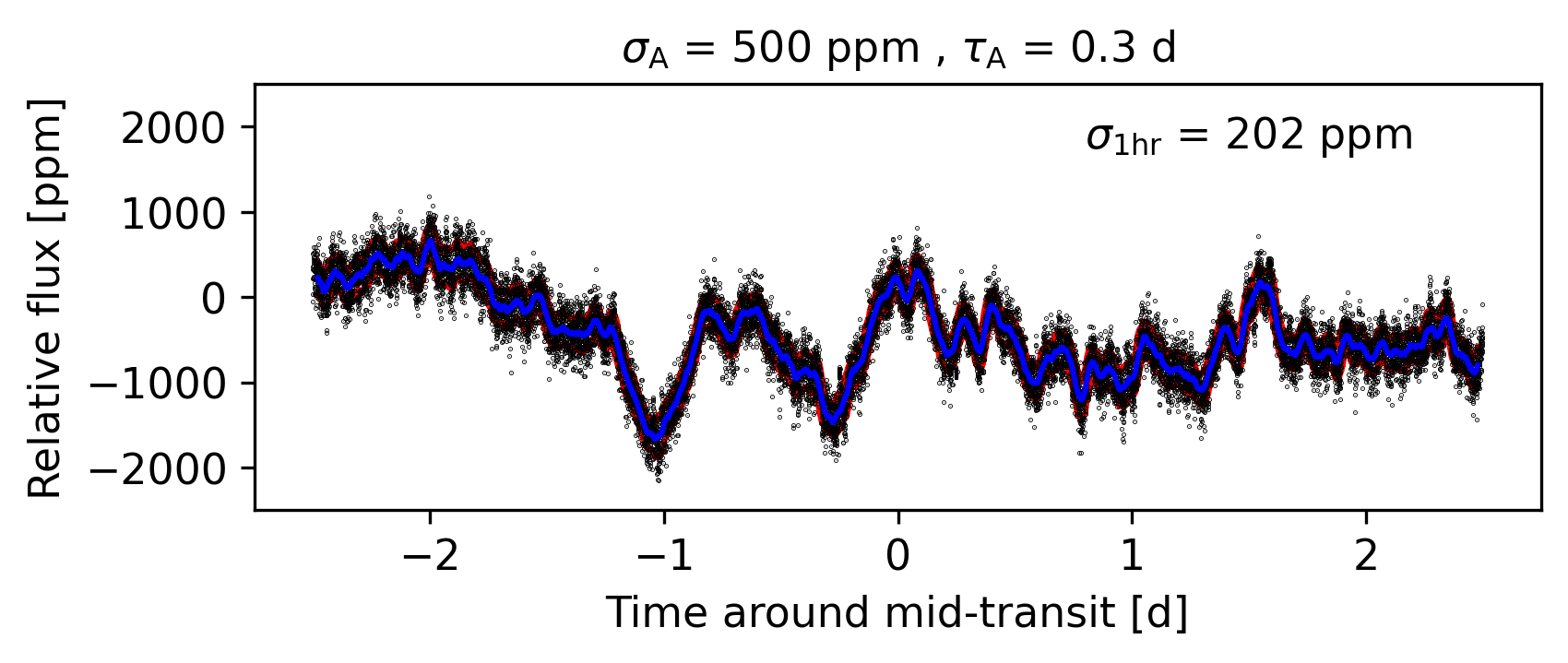}
    \caption{Same as Fig.~\ref{fig:lightcurves_1} but for an active star ($\sigma_{\rm A}=500$\,ppm, $\tau_{\rm A}=0.3$\,d). Note the vastly different range along the ordinate compared to Figs.~\ref{fig:lightcurves_1} and \ref{fig:lightcurves_2}.}
    \label{fig:lightcurves_3}
\end{figure*}

For each star, we generated 40 light curves with two transits of an Earth-like planet and an equal amount of light curves without transits. In each light curve we measured the sliding standard deviation in a 1\,hr window as a proxy for the combined activity on that timescale ($\sigma_{\rm 1\,hr}$). All simulations assumed an apparent stellar magnitude of $m_V=9$ and coverage by 24 PLATO cameras. Our measurements of $\sigma_{\rm 1\,hr}$ for the three benchmark stars are given in Table~\ref{tab:activity}. In Figs.~\ref{fig:lightcurves_1} - \ref{fig:lightcurves_3} we show some examples of the resulting light curves, each of which includes a transit of an Earth-like planet. In our preliminary analysis, we found that the transit detectability with {\tt TLS} upon detrending with {\tt W{\={o}}tan} depends sensitively on $\sigma_{\rm A}$ but weakly on $\tau_{\rm A}$. As we quantified in a limited injection-recovery test, shallow transits of Earth-sized planets are securely recovered around the quiet benchmark star (Fig.~\ref{fig:lightcurves_1}) but a large fraction of them gets lost in the stellar activity around even the moderately active benchmark star (Fig.~\ref{fig:lightcurves_2}) and they become entirely undetectable around the more active stars (Fig.~\ref{fig:lightcurves_3}). These findings illustrate that our estimates of PLATO planet yield, which we derived using a photometrically quiet star, must be seen as upper limits.

\begin{table}[h]
\caption{Stellar activity measurements from simulated {\tt PSLS} light curves.}
\def\arraystretch{1.1}
\label{tab:activity}
\centering
    \begin{tabular}{c|c|c}
    \hline
    $\sigma_{\rm A}$ & $\tau_{\rm A}$ & $\sigma_{\rm 1\,hr}$ \\
    \hline
         40\,ppm & 0.8\,d & 173\,ppm \\
        166\,ppm & 0.5\,d & 178\,ppm \\
        500\,ppm & 0.3\,d & 198\,ppm \\
    \hline\hline
    \end{tabular}\\
\tablefoot{The parameters $\sigma_{\rm A}$ and $\tau_{\rm A}$ are used in {\tt PSLS} to characterize the amplitude and timescale of stellar activity \citep{Samadi_2019}. For comparison, $\sigma_{\rm 1\,hr}$ is the mean standard deviation that we extracted in a running 1\,hr window from the light curves. For each combination of $\sigma_{\rm A}$ and $\tau_{\rm A}$ we simulated 40 light curves.}
\end{table}

\end{document}